\documentclass[reprint,amsmath,amssymb,superscriptaddress,longbibliography,aps,floatfix,nofootinbib]{revtex4-1}
\usepackage{graphicx}
\usepackage{bm,mathdots,dsfont}
\usepackage[colorlinks]{hyperref}
\usepackage[mathlines]{lineno}

\newcommand{\mb}[1]{\mathbf{#1}}
\newcommand{\T}{\mathsf{T}}

\newcommand{\mbh}[1]{\mathbf{\hat{#1}}}

\newcommand{\dyadic}[1]{\overline{\overline{#1}}}

\newcommand{\dt}{\mathrm{d}t}
\newcommand{\dV}{\mathrm{d}V}
\newcommand{\domega}{\mathrm{d}\omega}
\newcommand{\im}{\mathrm{i}}
\newcommand{\blangle}{\big\langle}
\newcommand{\brangle}{\big\rangle}
\newcommand{\uvec}[1]{\hat{\mathbf{#1}}}

\newcommand{\Ebf}{\mathbf{E}}

\newcommand{\Jbf}{\mathbf{J}}
\newcommand{\rbf}{\mathbf{r}}

\newcommand{\epsdbar}{\dyadic{\varepsilon}}

\newcommand{\intt}{\int_{-\infty}^\infty}
\newcommand{\intV}{\int_V}

\makeatletter
\newcommand{\supplementtitleformat}{\frontmatter@title@format}
\newcommand{\supplementauthorformat}{\normalfont\normalsize}
\newcommand{\supplementaffiliationformat}{\normalfont\frontmatter@affiliationfont}
\makeatother

\begin{abstract}\noindent
Kirchhoff’s law fundamentally relates thermal emission to absorption. For linear, static, reciprocal media, it equates the emissivity and absorptivity for each direction and frequency, while in nonreciprocal systems emission and absorption are equal when the bias is time-reversed. In time-varying media, however, temporal modulation breaks time-translation invariance, converts frequencies, and enables energy exchange with the modulation drive. As a result, a same-frequency relation between absorptivity and emissivity can no longer be expected. Here, we derive a generalized Kirchhoff’s law for linear time-varying Floquet media. We show that the emissivity at a given frequency equals a weighted sum of harmonic-resolved absorptivities of the adjoint system, with weights accounting for thermal occupation and photon-flux conversion. This relation has both practical and fundamental consequences. In practical terms, it allows emissivity to be calculated from absorption, simplifying the design of time-varying thermal emitters. More fundamentally, it reveals thermal radiation regimes inaccessible in static media. In particular, we identify time-varying structures that exhibit strong emission with negligible absorption at the same frequency for all directions, yielding a near-maximal violation of the conventional form of Kirchhoff’s law.\end{abstract}

\begin{document}
\title{A generalized Kirchhoff's law of thermal radiation for Floquet media }
\author{Sander A. Mann}
\affiliation{Institute of Physics, University of Amsterdam, Amsterdam, The Netherlands.}
\affiliation{Photonics Initiative, Advanced Science Research Center, City University of New York, New York, NY 10031, USA.}
\author{Dimitrios L. Sounas}
\affiliation{Department of Electrical and Computer Engineering, Wayne State University, Detroit, MI, USA.}
\author{Andrea Al\`u}
\affiliation{Photonics Initiative, Advanced Science Research Center, City University of New York, New York, NY 10031, USA.}
\affiliation{Physics Program, Graduate Center, City University of New York, New York, NY 10016, USA.}

\maketitle

\begin{figure*}[t!]
\centering
\includegraphics[width=\textwidth]{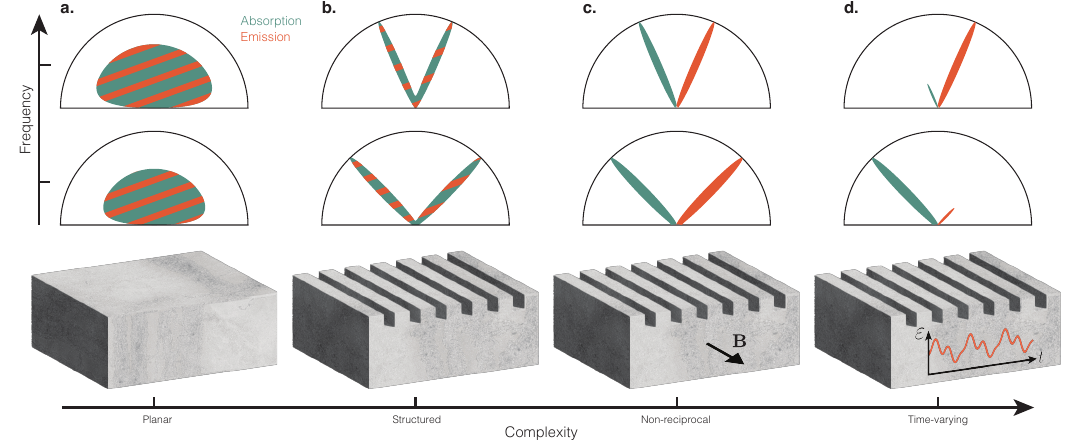}
\caption{{\bf Symmetry constraints on absorption and emission.} Thermal emission from media with progressively relaxed symmetry constraints: for a planar half-space ({\bf a}), the absorptivity and emissivity are simply dictated by its Fresnel reflection, yielding weak angular and frequency dependence as shown in the polar plots. For periodically structured materials ({\bf b}), absorption and emission can be highly directional, with strong frequency dependence and tailored polarization states, but absorptivity and emissivity are necessarily equal for each direction and frequency. This property no longer holds for materials with a magnetic bias ({\bf c}), which breaks reciprocity: now the emissivity and absorptivity can have different angular profiles. Yet, absorptivity and emissivity integrated over the full angular range must still be equal at each frequency. In time-varying media, time-translation symmetry is broken and it becomes possible to fully absorb from a specific direction at one frequency, but not emit at all at the same frequency ({\bf d}). }
\label{fig1}
\end{figure*}

\noindent
Kirchhoff's law is the cornerstone of our understanding of thermal radiation~\cite{KirchhoffEnglish}. In its most familiar form, it reads

\begin{equation}
\epsilon(\omega,\mbh{r},\mathbf{p}) = \alpha(\omega,-\mbh{r},\mathbf{p}^*).
\label{KirchhoffsLaw}
\end{equation}
 This expression, valid for linear, reciprocal, static bodies, equates emission at frequency $\omega$ into a direction and polarization state $(\mbh{r},\mb{p})$ with absorption into the time-reversed combination $(-\mbh{r},\mb{p}^*)$. Polished materials feature weak angular and frequency dispersion of their thermal emission (Fig.~\ref{fig1}a), while proper spatial patterning can shape the angular and frequency response to great effect (Fig.~\ref{fig1}b). However, in both cases, Kirchhoff's law constrains thermal emission frequency by frequency and for each angle of observation. This balance underlies a broad range of advances in thermal and light-generating photonics~\cite{vazquezlozano2024review,greffet2002coherent,dezoysa2012conversion,xu2021broadband,mohtashami2023metasurface,mou2024directiontunable,brar2015electronic,raman2014passive,lenert2014nanophotonic,xi2023camouflage}. 
 
The channel-by-channel equivalence between absorptivity and emissivity in Eq.~\ref{KirchhoffsLaw} is a consequence of Lorentz reciprocity \cite{snyder1998thermodynamic,guo2022adjoint}. Breaking reciprocity, for example by applying a magnetic bias, lifts this constraint: emission into a given channel no longer needs to equal absorption from that channel (see Fig.~\ref{fig1}c) \cite{zhao2019violation,zhao2020axion,shayegan2024broadband,yang2024nonreciprocal,zhang2025strong,guo2022adjoint}. Instead, the absorptivity for a given channel is now equal to the emissivity into the time-reversed channel of the adjoint system, i.e., with time-reversed bias. Interestingly, global thermal equilibrium still imposes a fundamental constraint: at each frequency, the emission and absorption integrated over all directions must be equal \cite{guo2022adjoint}.

Time-varying media offer more opportunities to mold the thermal emission properties. Temporal modulation breaks continuous time-translation symmetry, mixes frequencies, and enables energy exchange with the external drive, providing not only a route to magnet-free nonreciprocity, but also efficient frequency conversion and time-scattering phenomena~\cite{sounas2017nonreciprocity,moussa2023temporal}. Some of the new phenomena enabled by time-modulation in the context of scattering, emission and absorption have been recently demonstrated in temporally asymmetric antennas and space-time metasurfaces across the microwave, mid, and near-infrared frequency ranges \cite{hadad2016breaking,cardin2020surface,sisler2024electrically,efimov2026nonreciprocal}.Temporal modulation has indeed started to open new directions in thermal photonics, where the modulation has been predicted to modify coherence, correlations and heat transfer, and to enable photonic refrigeration and super-Planckian emission \cite{li2019adiabatic,buddhiraju2020refrigeration,ghanekar2022violation,ghanekar2022nonreciprocal,yu2023coherence,vazquez2023incandescent,biehs2023breakdown,tang2024modulating,yu2024time,liberal2025can}. These developments, however, expose a  gap in our fundamental understanding and modeling of thermal radiation in time-varying systems: when emission at one frequency can be generated by fluctuations at many frequencies, and absorption at the same frequency can dissipate through many harmonics, does a relation between them --- akin to Kirchhoff's law --- exist? A general relation between absorptivity and emissivity in time-varying (Floquet) thermal emitters is lacking, and its existence would not only unveil a better understanding of the underlying processes behind thermal emission in driven systems, but also offer a powerful tool to model and optimize time-varying thermal emission to overcome the limitations of static emitters.

Here, we introduce and demonstrate a generalized Kirchhoff’s law for Floquet systems, showing that the emissivity at a given frequency is equal to the weighted sum of harmonic-resolved absorptivities. This relation reveals thermal radiation regimes that are inaccessible in static media. In particular, we demonstrate optimal systems for which temporal modulation produces channel-integrated emissivity at a singly frequency that far exceeds the total absorptivity, beyond what is possible in time-invariant nonreciprocal systems (Fig.~\ref{fig1}d).

\begin{figure*}[th!]
\centering
\includegraphics[width=\textwidth]{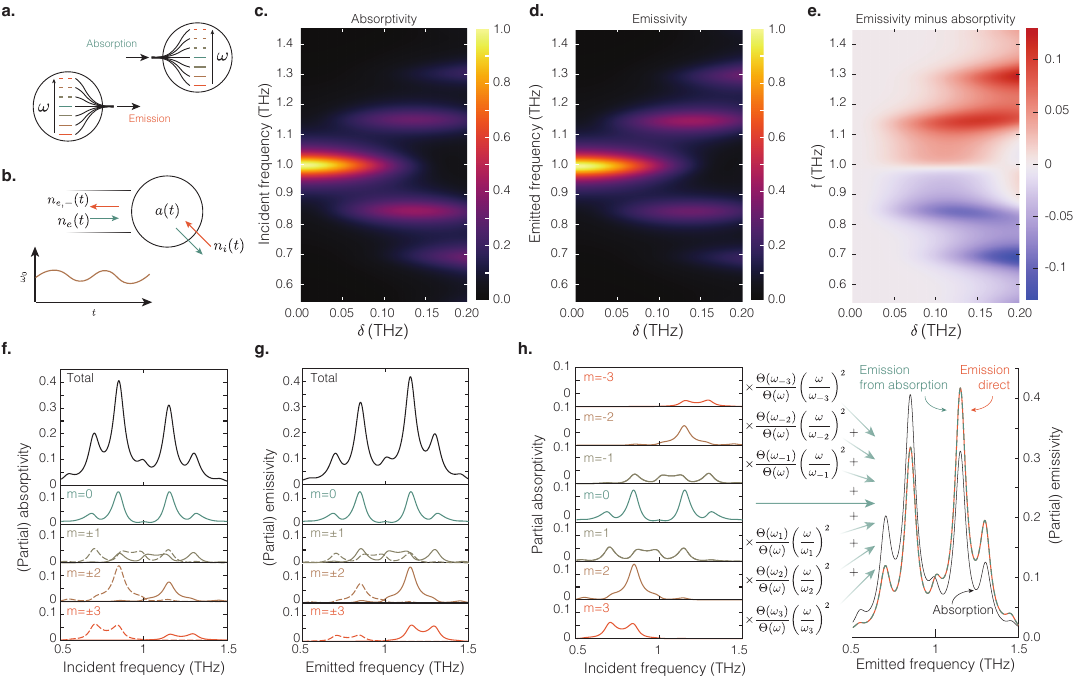}
\caption{{\bf Time-varying thermal emission from a modulated resonator. a.} Absorption and emission in time-varying systems: light incident at $\omega$ is distributed across harmonics and absorbed, while fluctuations at multiple frequencies are emitted at frequency $\omega$. {\bf b.} Schematic of an absorbing resonator with time-modulated center frequency. {\bf c,d.} Absorptivity ({\bf c}) and emissivity ({\bf d}) for a resonance with center frequency of 1 THz, modulated sinusoidally at 0.15 THz with increasing modulation depth on the horizontal axis. As the modulation depth increases, absorption and emission peaks at higher sidebands. {\bf e.} Difference between emissivity and absorptivity versus modulation depth. For frequencies above the unmodulated resonance frequency, emissivity exceeds absorptivity, while for frequencies below the emissivity is lower. {\bf f,g.} Partial absorptivities ({\bf f}) and emissivities ({\bf g}) at $\delta = 0.15$ THz up until the third harmonic. The positive (dashed) and negative (solid) partial absorptivities and emissivities are of similar shape but unequal in magnitude and displaced in frequency. {\bf h.} The total emissivity can be constructed from the partial absorptivities by multiplying them with the prefactor in the generalized Kirchhoff's law for time-varying media, Eq.~\ref{eq:genKlaw}. }
\label{fig2}
\end{figure*}

\section*{Floquet absorptivity and emissivity}
\noindent
Time-modulated systems require a more careful definition of absorption and emission than static systems, because frequency conversion allows energy injected at one frequency to be dissipated or emitted at another frequency.
%
%
As a result, in time-varying media, absorption and emission are sums over processes involving all harmonics instead of single frequency processes. Consider absorption: light incident at frequency $\omega$ can be converted to harmonics at $\omega_m = \omega + m \Omega$ before being absorbed, where $m\in \mathbb{Z}$ is the harmonic index and $\Omega$ the modulation frequency. The absorptivity at $\omega$ must be understood as the fraction of incident power at $\omega$ that is dissipated inside the structure \emph{at any frequency} (see Fig.~\ref{fig2}a). This naturally leads to a harmonic decomposition of the absorptivity:

\begin{equation}
\alpha(-\uvec{r},\uvec{p},\omega) = \sum_m \alpha_m(-\uvec{r},\uvec{p},\omega). \label{eq:absorptivity}
\end{equation}
$\alpha_m$ is the partial absorptivity in harmonic $m$, and can be derived from the expression for dissipation of electric fields $\mb{E}_m$ at harmonic $m$, $\alpha_m = (\omega_m/2) \intV \mb{E}^\dagger_m \epsdbar_\mathrm{I} \mb{E}_m \dV'$. Substituting the Green's function integral representation for $\mb{E}_m$ yields

\begin{multline}
    \alpha_m(-\uvec{r},\uvec{p},\omega) = \frac{16\pi^2c}{\omega\mu_0} \frac{\omega_m}{\omega} \times \\ \lim_{r\to\infty} r^2 \intV     \uvec{p}^\dagger \mb{G}_m^\dagger(\rbf',\rbf,\omega) \bm{\varepsilon}_\mathrm{I}(\rbf',\omega_m)  \mb{G}_m(\rbf',\rbf,\omega) \uvec{p} \,\dV'. \label{eq:absorptivity-harmonics}
\end{multline}
Here $ \mb{G}_m(\rbf',\rbf,\omega)$ is the Floquet Green's function for a source at $\rbf$ to a field at $\rbf'$ from $\omega$ to $\omega_m$ and $\bm{\varepsilon}_\mathrm{I}(\rbf',\omega_m)$ is the anti-Hermitian part of the permittivity tensor. 

The emissivity is defined analogously, but in reverse. We take it to be the emitted power at $\omega$ outside of the structure due to fluctuating currents $\emph{at any frequency}$ inside the structure, normalized to blackbody emission at the same temperature (see Fig.~\ref{fig2}a). Because fluctuating currents at $\omega_m$ inside the structure can be converted by the modulation into radiation at $\omega$, the emissivity decomposes as 

\begin{equation}
\epsilon(\uvec{r},\uvec{p},\omega)  =  \sum_m \frac{\Theta(\omega_m,T)}{\Theta(\omega,T)} \epsilon_m (\uvec{r},\uvec{p},\omega) . \label{eq:emissivitymatrix}
\end{equation}
The ratio $\Theta(\omega_m,T)/\Theta(\omega,T)$ accounts for different spectral densities of thermal fluctuations at different harmonics, where $\Theta(\omega,T) = \hbar \omega/(e^{\hbar \omega/k_B T}-1)$ is the mean thermal energy of a harmonic oscillator at $\omega$ and temperature $T$. $\epsilon_m$ is the partial emissivity due to sources with frequency $\omega_m$ for power emitted at the observation frequency $\omega$, given by

\begin{multline}
    \epsilon_m(\uvec{r},\hat{\mathbf{p}},\omega) = \frac{16\pi^2 c}{\omega\mu_0} \frac{\omega_m}{\omega} 
    \lim_{r\to\infty} r^2 \intV \uvec{p}^\dagger \mb{G}_{-m}(\rbf,\rbf',\omega_m)\times \\
    \bm{\varepsilon}_\mathrm{I}(\rbf',\omega_m) \mb{G}^\dagger_{-m}(\rbf,\rbf',\omega_m) \uvec{p} \,\dV'.
    \label{eq:emissivity-harmonics}
\end{multline}
Details of the derivation of Eqs.~\ref{eq:absorptivity}-\ref{eq:emissivity-harmonics} can be found in the Methods section and Supplementary Materials. More generally, the absorption and emission are characterized by absorptivity and emissivity operators, which capture correlations between  directions and frequencies. The derivation of these operators can also be found in the Supplementary Materials.

In time-invariant media $\alpha$ and $\epsilon$ can be obtained directly from the diagonal elements of  $1-\mathbf{S}^\dagger \mathbf{S}$ and $1-\mathbf{SS}^\dagger$, respectively, where $\mathbf{S}$ is the scattering matrix~\cite{miller2017universal}. This approach can be extended to time-varying media in a photon basis by  assuming a constant thermal occupation in frequency~\cite{ghanekar2022nonreciprocal,efimov2026nonreciprocal}, which is a valid assumption for low modulation frequencies. However, if one is not constrained to slow modulation it is necessary to calculate  $\epsilon$ from fluctuation currents directly, considering that an emitted photon at a given frequency is generally the result of fluctuation currents over multiple harmonics with different occupation factors. Similarly, for the absorbed power, it is not enough to know the total number of photons that enter the medium, but through which harmonics each of these photons is absorbed. These questions can only be addressed through the fluctuation current approach presented here.

The similarity of the partial absorptivity Eq.~\ref{eq:absorptivity-harmonics} and emissivity Eq.~\ref{eq:emissivity-harmonics}  suggests that a Kirchhoff-like relation may survive even in time-modulated systems. As shown rigorously in the Methods and Supplementary Materials Sections 1 and 2, this intuition can indeed be made exact by using the reciprocity condition 

\begin{equation}
\frac{\mb{G}^\T_{m}(\mb{r},\mb{r}',\omega)}{\omega_m} = \frac{\mb{G}_{-m}^{\mathcal{T}}(\mb{r}',\mb{r},\omega_m)}{\omega}
\end{equation}
on the Green's function in time-varying media \cite{asadchy2020tutorial}. By doing so, we find a generalized Kirchhoff's law for Floquet media, which is the main result of this work:

\begin{equation}
\epsilon(\uvec{r},\uvec{p},\omega) = \sum_m \frac{\Theta(\omega_m,T)}{\Theta(\omega,T)} \biggl( \frac{\omega}{\omega_m} \biggr) ^2 \alpha_{m}^\mathcal{T}(-\uvec{r},\uvec{p}^*,\omega).
\label{eq:genKlaw}
\end{equation}
This equation expresses the emissivity at emitted frequency $\omega$ in terms of the partial absorptivities of the adjoint system at all harmonics, $\alpha_m^\mathcal{T}$ with $\mathcal{T}$ indicating the adjoint system (with time-reversed bias). The ratio of the thermal spectral density $\Theta$ accounts for the difference in fluctuations at the different harmonics, while the factor $(\omega/\omega_m)^2$ originates from the Manley-Rowe conversion between power and photon flux. This relation is derived for linear, periodically driven systems with stationary thermal reservoirs and dissipation and modulated lossless regions.

\section*{Emission from a time-varying resonator}

\noindent
Armed with these expressions for the absorptivity and emissivity in time-varying media, we can explore how temporal modulation affects them. We first consider a canonical problem: a resonator with harmonically modulated center frequency $\omega_0(t)=\omega_s + \delta \cos(\Omega t)$ (Fig.~\ref{fig2}b). We take the static center frequency $\omega_s/2\pi$ to be 1 THz and the radiation and absorption decay rates $\gamma_e$ and $\gamma_i$ to be $\gamma_e=\gamma_i= 0.02 \omega_s$, implying that the resonator is critically coupled and achieves unity absorptivity/emissivity at resonance. We modulate the center frequency with frequency $\Omega/2\pi = 0.15$ THz, and vary the modulation depth $\delta/\omega_s$ from 0 to 20\%. The resulting absorptivity is shown in Fig.~\ref{fig2}c. At $\delta=0$ we observe strong absorption at the resonance frequency, as expected for a critically coupled resonator. With increasing modulation depth, however, absorption at the center frequency decreases while it increases in the first positive and negative harmonic. For even stronger modulation depths, absorption is strongest in the next harmonics, and so forth. The emissivity, shown in Fig.~\ref{fig2}d, at first glance looks similar, but when we subtract the absorptivity from the emissivity clear differences emerge: Fig.~\ref{fig2}e shows that the emissivity is significantly larger than the absorptivity for frequencies above the unmodulated resonance frequency, but lower below it. This is a consequence of the larger spectral density of fluctuations at lower frequencies and the Manley-Rowe frequency factor, and can be used to achieve photonic refrigeration \cite{buddhiraju2020refrigeration}. 

The absorptivity and emissivity in different harmonics are visualized directly in Figs.~\ref{fig2}f,g for a relative modulation depth of 10\%. We observe that the power absorbed at positive and negative harmonics is similar, but displaced in frequency and different in magnitude due to up- or down conversion before absorption. The partial absorptivity gradually decreases for larger harmonics and is essentially absent above the third harmonic. The partial emissivities exhibit similar behavior, except that the emitted power from negative harmonics is now higher due to the reversed role of up- and down conversion in addition to the spectral density prefactor. 

To visualize the generalized Kirchhoff's law for the system in Fig.~\ref{fig2}, we show how the partial harmonics multiplied by the prefactors add up to the emissivity in Fig.~\ref{fig2}h. Since there is only a single harmonic modulation tone, the system is time-reversal symmetric and the emissivity is simply given by a sum of the partial absorptivities scaled by the prefactor. Fig.~\ref{fig2}h shows how these scaled partial harmonics add up to an emissivity curve identical to the directly calculated emissivity, validating the generalized Kirchhoff's law for time-varying media Eq.~\ref{eq:genKlaw}.

\begin{figure}[t!]
\centering
\includegraphics[width=0.45\textwidth]{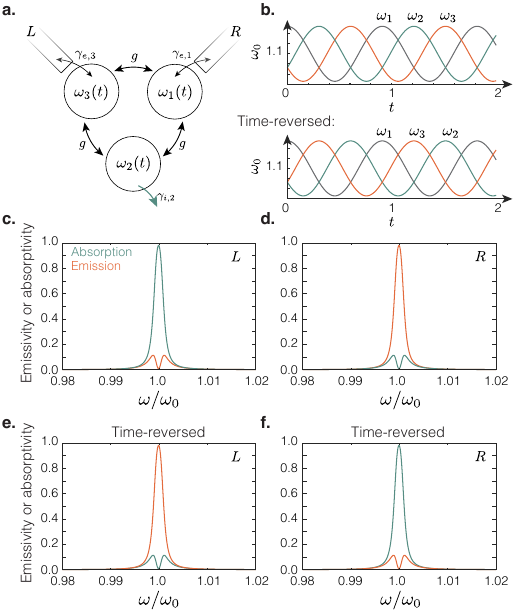}
\caption{{\bf A nonreciprocal thermal emitter through time-modulation.} {\bf a}. Schematic of a time-modulated ring of three resonators coupled to two ports, with one of them absorbing. {\bf b.} Each resonance frequency is modulated at frequency $\Omega = 0.05 \omega_0$ but with a $2\pi/3$ phase difference, inducing angular momentum bias. Under time-reversal the sign of the phases changes (bottom). {\bf c.} In the left port, absorption peaks at $\omega_0$ but the emissivity dips, demonstrating a large difference between absorptivity and emissivity for the same frequency at the same port. {\bf d.} In the right port, we observe the reverse behavior: emission peaks while absorption is absent. This is consistent with the behavior of a nonreciprocal time-invariant thermal emitter, such as a magnetically biased one. {\bf e,f.} Under time-reversal, the behavior in the left ({\bf e}) and right ({\bf f}) ports reverses, consistent with the behavior of a magnetically-biased thermal emitter.  }
\label{fig3}
\end{figure}

The model system considered so far is time-reversal symmetric, so the unequal emissivity and absorptivity do not stem from static nonreciprocity as in recent work~\cite{shayegan2023direct,shayegan2024broadband,zhang2025strong,yang2024nonreciprocal}, but entirely from work done by the modulation and the sampling of thermal fluctuations at other harmonics with different occupation factors. Larger differences between emissivity and absorptivity can be expected by additionally breaking reciprocity in optimized structures.  

\section*{Nonreciprocal thermal radiation}\noindent
Using time-modulation, we can break reciprocity by synthesizing an effective angular momentum bias~\cite{sounas2017nonreciprocity,estep2014,yang2024nonreciprocal}. Consider a loop of three resonators, each with a resonance frequency that is modulated in time (Fig.~\ref{fig3}a) with a $2\pi/3$ phase shift compared to the next one, as shown in Fig.~\ref{fig3}b. This modulation scheme induces a rotating modulation pattern, whose handedness breaks time-reversal symmetry and reciprocity. The result is that such a device can operate as a circulator, where certain frequency signals only travel counterclockwise (or clockwise for time-reversed phases, Fig.~\ref{fig3}b, bottom) \cite{estep2014}, similar to a magnetically biased circulator. 

By connecting a left (L) and a right (R) port to resonators 3 and 1 respectively, and by adding absorption losses to resonator 2, we can break the conventional Kirchhoff's law. Inspecting absorptivity and emissivity in the left port in Fig.~\ref{fig3}c, we observe strong absorption and very little emissivity, demonstrating a large violation of Eq.~\ref{KirchhoffsLaw}. By contrast, when we inspect the right port, we observe the opposite behavior: the emissivity is large but the absorptivity is small at the same resonance frequency. In this case the sideband contributions to absorption and emission are small, so Eq.~\ref{eq:genKlaw} reduces to $\epsilon(\uvec{r},\uvec{p},\omega) \approx \alpha^\mathcal{T}(-\uvec{r},\uvec{p}^*,\omega)$. As such, this system is analogous to a magnetically biased static nonreciprocal emitter. The emissivity should be equal to the absorptivity in the time-reversed system, which is verified in Figs.~\ref{fig3}e,f. Importantly, in this weak modulation limit the emissivity and absorptivity summed over both ports must be equal~\cite{guo2022adjoint}. In magnetically biased emitters, this indeed \emph{must} be the case: while the emissivity and absorptivity for a specific direction may be dramatically different, in order to satisfy the second law of thermodynamics, the emissivity and absorptivity integrated \emph{over all angles} must be equal~\cite{guo2022adjoint,yang2024nonreciprocal}.

\section*{Maximal violation of Kirchhoff's law}\noindent
While the example of Fig.~\ref{fig3} demonstrates how time-modulation can be used to violate the conventional Kirchhoff's law, this perturbative scheme still complies with the thermodynamic constraints of passive nonreciprocal systems. Yet, time-varying media break time-translation invariance, implying that energy exchange between the system and the modulation network is possible, further relaxing the relation between absorption and emission. We can therefore expect the possibility of structures that emit strongly at a given frequency, while not absorbing in any channel for the same frequency, a phenomenon that would represent a much stronger violation of the standard Kirchhoff's law and goes beyond what can be achieved in time-invariant nonreciprocal emitters. 

\begin{figure}[t!]
\centering
\includegraphics[width=0.45\textwidth]{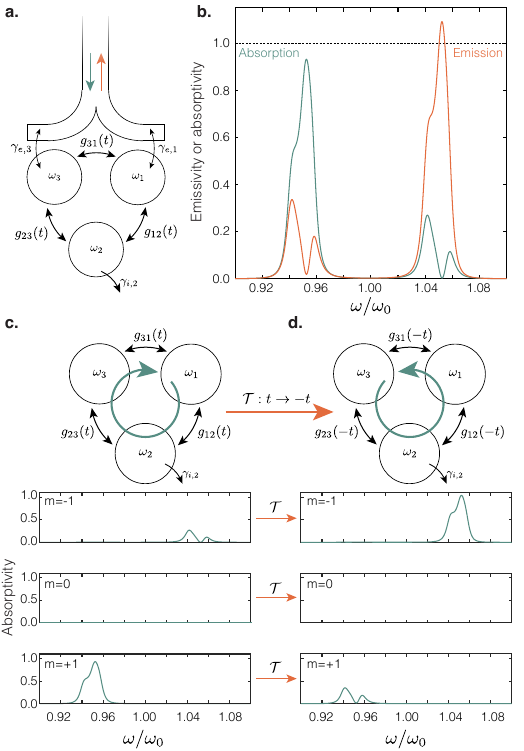}
\caption{{\bf Maximal violation of Kirchhoff's law.} {\bf a}. Schematic of a loop of detuned resonators with couplings modulated in time at the difference frequencies between the connected resonances, thereby efficiently up- and down-converting around the loop with a handedness defined by the modulation phases. Two of the resonances ($\omega_1$ and $\omega_3$) are coupled to the same input port, while the third resonance $\omega_2$ absorbs. {\bf b.} Absorptivity and emissivity at the port, demonstrating near-maximal violation of the standard Kirchhoff's law: strong absorption without emission and vice-versa. {\bf c.} The absorptivity of this structure can be understood via its harmonics: at the fundamental frequency absorption is absent, since at $\omega_2$ there is no coupling between the port and the resonance. Strong absorption occurs at the $+1$ harmonic, when light is incident resonant with $\omega_1$ and upconverted to be absorbed. {\bf d.} Under time-reversal, the loop changes direction, and we observe strong absorption in the $-1$ harmonic. This is in agreement with the generalized Kirchhoff's law Eq.~\ref{eq:genKlaw}, which states that the emissivity is related to the absorptivity in the time-reversed system.}
\label{fig4}
\end{figure}

To demonstrate extreme violation of Kirchhoff's law, we consider an emitter with a single input/output port. In time-invariant systems, independent of reciprocity, the emissivity and absorptivity of a one-port system \emph{must} be equal because of thermodynamic equilibrium. However, as we show in the following, in time-varying systems this is not the case, and our generalized Kirchhoff's law Eq.~\ref{eq:genKlaw} enables a straightforward design of such systems. We again consider a loop of three resonators, with relative un-modulated resonances at 0.96, 1, and 1.04 $\omega_0$ (see Fig.~\ref{fig4}a). They are coupled to each other with coupling rates $g(t)$ modulated by the difference frequency of the two resonators they connect, and the phase of the coupling modulation differ by $2\pi/3$ again. Due to the modulated coupling, photons can be up- and down-converted and move from one cavity to another, despite the frequency difference. The phase differences again induce a specified handedness to the hopping, resulting in a circulator-type motion: photons from the first resonance are up-converted to the second one, from the second to the third, and then from the third are down-converted to the first.

We again set only the second resonance to be absorbing with rate $\gamma_{i,2}$, while both the first and third cavity are coupled to the single input/output port with rates $\gamma_{e,1}=\gamma_{e,3}$. By setting the magnitude of the coupling rate $|g| = \gamma_{i,2} = \gamma_{e,1} = \gamma_{e,3} = 4 \times 10^{-3} \omega_0$, we observe spectral features for the absorptivity and emissivity that are drastically different in Fig.~\ref{fig4}b: at the frequency for which the emissivity peaks, the absorptivity nearly vanishes, and vice versa, due to the handedness of the harmonic loop. This maximal violation of Kirchhoff's law in a system with a single input/output port, yielding absorptivity (emissivity) near-unity and negligible emissivity (absorptivity), can only occur in time-varying systems where frequency conversion and energy exchanges with the modulation network play a significant role.

To gain further insights into the origin of this remarkable phenomenon, we can visualize the partial absorptivity at each harmonic, in the same vein as our generalized Kirchhoff's law Eq. 7. Despite absorption in the second resonance, incident light with frequency close to $\omega_2$ is not absorbed, since the second resonance is not coupled to the input/output port. Indeed, Fig.~\ref{fig4}c shows that absorption at the fundamental frequency is absent for any frequency. Instead, incident light with frequency close to $\omega_1$ couples into the first resonance, is then up-converted to $\omega_2$ where it can be absorbed: the $m=+1$ harmonic in Fig.~\ref{fig4}c shows a clear absorption peak. Incident light with frequency close to $\omega_3$ also enters the structure, but in the third (lossless) resonator. It is subsequently down-converted to $\omega_1$, where it primarily leaks out into the port again. This can be observed in the $m=-1$ harmonic, where absorption is weak. The sum of the partial absorptivities yields the total absorptivity spectrum shown in Fig.~\ref{fig4}b.

For emission, fluctuations generate photons in the second resonance at $\omega_2$, which are upconverted to $\omega_3$ and emitted. This process can also be observed in Fig.~\ref{fig4}b: the emissivity peaks at $\omega_3$, while it is low at $\omega_1$. Interestingly, the emissivity is super-Planckian (exceeds unity), due to the prefactors in Eq.~\ref{eq:genKlaw}: as emitted photons are upconverted from a frequency with higher spectral density of fluctuations than at the emitted frequency, thereby resulting in emission above that of a blackbody \cite{xiao2022} and photonic refrigeration \cite{buddhiraju2020refrigeration}. Note that the excess relative to the blackbody limit is enabled by work supplied by the modulation drive, and is not possible in passive equilibrium systems.

Our generalized Kirchhoff's law relates the emissivity to the absorptivity in the time-reversed system. Indeed, by reversing the temporal modulation scheme in the geometry of Fig. 4, the partial absorptivities are reversed, as shown in Fig.~\ref{fig4}d. Now, absorption is highest in the $m=-1$ harmonic, for frequencies incident at $\omega_3$ which are subsequently down converted to $\omega_2$ and absorbed. Scaling these partial absorptivities with the prefactor in Eq.~\ref{eq:genKlaw} and summing them yields the emissivity shown by the dashed line in Fig.~\ref{fig4}b, confirming the generalized Kirchhoff's law also for structures with broken time-reversal symmetry. 

\begin{figure}[t!]
\centering
\includegraphics[width=0.48\textwidth]{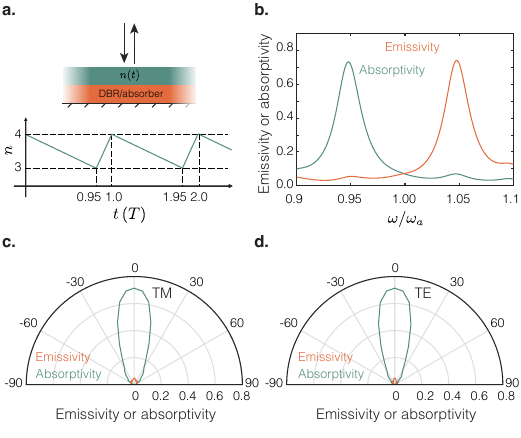}
\caption{{\bf Maximal violation of Kirchhoff's law in a full-wave structure.} {\bf a}. A time-modulated layer is positioned on top of a narrowband absorber, formed by an absorbing layer in a distributed Bragg cavity absorbing at $\omega_a$. The top layer is modulated with a sawtooth profile with period $2\pi/T = \omega_a/20 $. {\bf b.} The absorptivity and emissivity at normal incidence, demonstrating strong violation of the time-invariant Kirchhoff's law. {\bf c,d.} The angle dependence of the absorptivity and emissivity at $\omega = 0.95 \omega_a$ for the TM ({\bf c}) and TE ({\bf d})  polarization, showing that the violation persists for all angles and both polarizations. }
\label{fig5}
\end{figure}

Strong violation of the conventional Kirchhoff's law can  also expected in time-modulated full-wave systems. As an example, consider a narrowband absorber, for instance realized through a thin absorbing layer embedded in a distributed Bragg reflector with resonance centered at $\omega_a$, placed on top of a perfect electric conductor, and covered with a lossless material with a time-varying permittivity (Fig.~\ref{fig5}a). Rather than a single-tone harmonic modulation, the permittivity follows a sawtooth modulation that breaks time-reversal symmetry and upconverts radiation incident upon it from any direction of incidence \cite{ramaccia2020electromagnetic}. As a result, the incident light is upconverted and then absorbed. The emitted light by fluctuating currents in the absorbing layer is also upconverted before being radiated away. 

Since both the emmitted and absorbed light are upconverted, they peak at different frequencies. This can be observed in Fig.~\ref{fig5}b, which shows full-wave simulations of the absorptivity and emissivity of this structure, with the emissivity calculated through our generalized Kirchhoff's law Eq.~\ref{eq:genKlaw}, and the partial absorptivities of the time-reversed system. The absorptivity and emissivity at normal incidence are spectrally separated by a frequency twice the modulation frequency, as expected from the upconversion action of the modulation. It is important to emphasize that in this structure the large difference between emissivity and absorptivity persists for all angles of incidence, as shown in the polar plots in Fig.~\ref{fig5}c,d. On the contrary, in time-invariant nonreciprocal structures the integrated emissivities and absorptivities over all angles are necessarily equal~\cite{guo2022adjoint,yang2024nonreciprocal}. 

\section*{Validity and implications}\noindent
The central consequence of our generalized Kirchhoff's law is that, because emission is generated by fluctuations at multiple harmonics, the correct relation should be established with each harmonic in the adjoint system weighted by the thermal occupation and photon-flux conversion factor. This distinction is clearly visible in Fig.~\ref{fig2}, where unequal same-frequency emission and absorption arises in a time-reversal symmetric system, and the difference between absorptivity and emissivity becomes much stronger in the nonreciprocal examples of Figs.~\ref{fig4} and \ref{fig5}.

The law presented here applies to periodically driven systems coupled to stationary thermal reservoirs. This means that the modulation only acts on the lossless, reactive part of the system, while the dissipative materials and their associated thermal reservoirs remain time invariant. One realization is a structure in which time-invariant lossy materials are embedded in a time-varying lossless medium. Such an approach is common when modeling time-varying emitters, because under these conditions the standard fluctuation-dissipation theorem is still rigorously valid in the dissipative regions. If the dissipative regions themselves are modulated, the associated noise currents become nonstationary and correlated across Floquet harmonics, requiring a generalized fluctuation-dissipation relation or a microscopic time-dependent reservoir model~\cite{yu2024time,vazquez2023incandescent,horsley2025mqed}.

In our derivation, we have focused on the thermal contribution to emission, and have not included zero-point fluctuations. At high frequencies and/or low temperatures, quantum commutation constraints may become important, requiring extra care~\cite{vazquez2023incandescent}. Under these assumptions, however, Eq.~\ref{eq:genKlaw} provides a powerful and practical design rule: the emissivity can be computed from absorption in the adjoint system, rather than from direct fluctuating source simulations. This provides a route to designing time-varying emitters with emission and absorption channels separated in frequency, direction and polarization, with possible applications in thermal routing, photonic refrigeration, and engineered radiative noise environments.

\section*{Acknowledgements}\noindent
We acknowledge Loubnan Abou Hamdan for feedback on the manuscript. This research is based upon work supported in part by the Air Force Office of Scientific Research SBIR program under the technical guidance of Dr. A. Nachman.

\clearpage

\section*{Methods}
\subsection*{Langevin coupled-mode theory}\noindent
To elucidate the relationship between absorption and emission of thermal radiation in time-varying systems, in Figs.~\ref{fig2}-\ref{fig4} we first consider resonant systems well described by a limited set of resonant modes interacting with input and output channels. This assumption allows us to model the response with coupled-mode theory \cite{suh2004temporal}, featuring a collection of coupled resonances at frequencies $\omega$ with complex amplitudes $\mb{a}$, normalized such that their absolute value squared is the stored energy in each resonance.

In the context of thermal radiation, coupled-mode theory lends itself well to study emission from strongly nonlinear systems \cite{khandekar2015radiative} as well as from time-varying systems \cite{zhu2013temporal,buddhiraju2020refrigeration}. In order to capture thermal radiation from such system, we describe its dynamics through the Langevin equation 

\begin{equation}
\frac{d}{dt}\mb{a} = -i(\mb{H}(t) - i\mb{\Gamma}_e - i\mb{\Gamma}_i)\mb{a} + \mb{K}^\T \mb{n}_e^+ +  \mb{L}^\T\mb{n}_i,
\label{eommod}
\end{equation}
where we assume $\exp(-i\omega t)$ time convention and $\mb{\Gamma}_e$, and $\mb{\Gamma}_i$ are Hermitian matrices describing decay due to external radiation and internal dissipation respectively, including (dissipative) coupling in the off-diagonal elements. The resonances are coupled to external baths (via the input channels) and to internal reservoirs (via dissipation) through the matrices $\mb{K}$ and $\mb{L}$. The  stochastic Langevin sources $\mb{n}_e$ and $\mb{n}_i$ represent the excitation of the resonances due to incident thermal radiation and fluctuating currents inside the media, respectively. The temporal variation of the structure is captured in $\mb{H}$.

The thermal emission emanating from the structure is given by

\begin{equation}
\mb{n}_e^-= \mb{C}\mb{n}_e^+ + \mb{D} \mb{a},
\label{reflected}
\end{equation}
where $\mb{C}$ is a direct reflection matrix that captures scattering between the ports in the absence of the resonances and $\mb{D}$ is a matrix describing the coupling between the resonators and the output channels. These  matrices are not all independent: they obey the identities $\mb{CD}^*=-\mb{K}$, $2\mb{\Gamma}_e = \mb{K}^\dagger \mb{K}$, and $   2\mb{\Gamma}_i = \mb{L}^\dagger \mb{L}$ \cite{suh2004temporal,mann2019nonreciprocal}.

We assume a periodic drive with modulation period $T = 2\pi/\Omega$. Hence, we can move to the frequency domain and expand Eq.~\ref{eommod} into a system of equations for all harmonics. Solving this system for the resonance amplitudes (see Supplementary Materials for full derivation), we find:

\begin{equation}
\mbh{a} = -\mbh{G}(\mbh{K}^\T\mbh{n}_e +  \mbh{L}^\T\mbh{n}_i)
\label{amplitudes}
\end{equation}
Here, $\mbh{a}=(\cdots,\mb{a}_{1}, \mb{a}_{0}, \mb{a}_{-1},  \cdots)^\T$ is a vector containing the amplitudes of all harmonics, and the matrices contain the evolution of the harmonics along the diagonal and coupling between them due to modulation on off-diagonal submatrices. The matrix $\mbh{G} = (i\mbh{H} - i\mbh{\Omega} -\mbh{\Gamma}_e -\mbh{\Gamma})_i^{-1}$ is the Green's function with $\mbh{\Omega} = \text{diag}(\omega_n \mb{1}_{k\times k} )$ a block diagonal matrix with the frequencies of the harmonics $\omega_n$ and each block size given by the number of resonances. The stochastic sources satisfy $\langle n_{e,p}(\omega) n_{e,q}(\omega') \rangle= \langle n_{i,p}(\omega) n_{i,q}(\omega') \rangle = \Theta(\omega,T) \delta_{pq}\delta(\omega-\omega')$, where $\Theta(\omega,T) = \hbar \omega/(e^{\hbar \omega/k_B T}-1)$ is the mean energy of the classical oscillator.

 We define the emissivity as the emitted power at $\omega$ outside of the structure due to fluctuating currents $\emph{at any frequency}$ inside the structure, relative to blackbody emission (see Fig.~\ref{fig2}a). To find expressions for the emissivity and absorptivity, we therefore set $\mb{n}_e$ or $\mb{n}_i$ to zero respectively, and calculate the ensemble averaged absorption $\langle 2 \mb{a} \Gamma_i \mb{a}_-^\dagger \rangle$ and emission $\langle \mb{n}_{e,-} \mb{n}_{e,-}^\dagger \rangle$ normalized to the blackbody spectral density $\Theta (\omega,T)$, which yields

\begin{equation}
\boldsymbol{\alpha} = \sum_m \frac{\omega}{\omega_m} (\mbh{D}^* \mbh{G}^\dagger \mbh{K}_i^\T)_{0n} (\mbh{K}_i^*\mbh{G}\mbh{D}^\T)_{n0} = \sum_n \frac{\omega}{\omega_m} \boldsymbol{\alpha}_m \label{absorptivitymatrix}
\end{equation}
\begin{multline}
\boldsymbol{\epsilon} = \sum_m \frac{\Theta(\omega_m,T)}{\Theta(\omega,T)}  \frac{\omega_m}{\omega}  (\mbh{D} \mbh{G} \mbh{K}_i^\dagger)_{0n} (\mbh{K}_i \mbh{G}^\dagger \mbh{D}^\dagger)_{n0} \\
= \sum_m \frac{\Theta(\omega_m,T)}{\Theta(\omega,T)} \boldsymbol{\epsilon}_m \label{emissivitymatrix}
\end{multline}
Here $\boldsymbol{\alpha}$ and $\boldsymbol{\epsilon}$ are absorptivity and emissivity matrices, whose diagonal elements contain the absorptivity and emissivity of the corresponding port for a normalized input vector, while the off-diagonal elements contain the correlation between different ports \cite{miller2017universal}. $\boldsymbol{\alpha}_m$ and $\boldsymbol{\epsilon}_m$ are the partial absorptivity and emissivity matrices in harmonic $m$ for light incident/emitted at the fundamental $\omega$, and $\omega_m = \omega + m\Omega$. From these expressions the generalized Kirchhoff's law, Eq.~\ref{eq:genKlaw}, can be derived.

The spectra in Figs.~\ref{fig2}-\ref{fig4} were computed by solving the frequency-domain Floquet CMT equations described in this subsection for the corresponding resonator networks. For each incident or emitted frequency, the Floquet system was truncated to a finite set of harmonics and the resulting linear system was solved for the steady-state harmonic amplitudes. The absorptivities and emissivities were calculated using Eqs.~\ref{absorptivitymatrix} and \ref{emissivitymatrix}.

\subsection*{Full-wave derivation}\noindent
In contrast to Figs.~\ref{fig2}-\ref{fig4}, developed using the Floquet extension of temporal coupled-mode theory described in the previous subsection in the Methods,  Fig.~\ref{fig5} is a full-wave result. In Supplementary Materials Section 2 we show that Eq.~\ref{eq:genKlaw} holds in the full-wave limit, and can be derived directly from Maxwell's equations. A course outline of the derivation is provided in the main text. We will briefly elaborate on the derivation here, and refer the reader to Supplementary Materials Section 2 for the full derivation. 

The electric field $\mb{E}_{mp}$ at harmonic $m$ due to currents at harmonic $p$ is given by
\begin{equation}
\mb{E}_{mp}(\mb{r}) =\int_{V'} \mb{G}_{mp}(\mb{r},\omega_m,\mb{r}',\omega_p) \mb{J}(\mb{r}',\omega_p) \,\dV'
\label{Efromj}
\end{equation}
This is an extension of the standard expression for radiation from a current distribution, where the multi-frequency Green's function $\mb{G}_{mp}(\mb{r},\omega_m,\mb{r}',\omega_p)$ gives the field at frequency $\omega_m = \omega + m \Omega$ and position $\mb{r}$ due to currents at $\omega_p = \omega + p \Omega$ in the volume $V'$. 

Applying Eq.~\ref{Efromj} for a source at a large distance from the object and following the expression for the dissipated power from Poynting's theorem,
we find an expression for the absorbed power as given in Eq.~\ref{eq:absorptivity-harmonics}. The thermal radiation is proportional to the thermal ensemble average $\langle \mb{E}(\mb{r},\omega)\mb{E}^\dagger(\mb{r},\omega)\rangle$ \cite{rytov1989principles}. Replacing the electric field in this expression from Eq.~\ref{Efromj} with $\Jbf$ equal to the fluctuation currents inside the object we find the radiated power
as
\begin{multline}
    W_\mathrm{rad}(\hat{\mathbf{r}},\hat{\mathbf{p}},\omega) = \frac{4}{\eta_0} \sum_n \lim_{r\to\infty} r^2 \intV \omega_m \Theta(|\omega_n|)\times \\ \uvec{p}^\dagger \mb{G}_{-n}(\rbf,\rbf',\omega_n) \bm{\varepsilon}_\mathrm{I}(\rbf',\omega_n) \mb{G}^\dagger_{-n}(\rbf,\rbf',\omega_n) \uvec{p} \,\dV',
\end{multline}
where we have used the fluctuation-dissipation theorem, $\big\langle \Jbf(\rbf,\omega) \Jbf^\dagger(\rbf',\omega') \big\rangle = 4\pi \omega \Theta(\omega) \bm{\varepsilon}_\mathrm{I}(\rbf,\omega)  \delta(\rbf-\rbf') \delta(\omega - \omega')$, 
and $\mb{G}_{-m}(\rbf,\rbf',\omega_m)$ is shorter notation for $\mb{G}_{0m}(\mb{r},\omega,\mb{r}',\omega_m)$. Dividing by the black-body spectrum, we find the partial emissivity in Eq.~\ref{eq:emissivity-harmonics}.

The expressions for the partial emissivities and absorptivities are remarkably similar, except for the Green's functions. Just as in the proof of the static Kirchhoff's law, however, we may follow a similar approach and invoke generalized reciprocity. For time-invariant systems the Green's function adheres to the reciprocity condition $\mb{G}^\T(\mb{r},\mb{r}') = \mb{G}^{\mathcal{T}}(\mb{r}',\mb{r})$, where the superscript $\mathcal{T}$ again indicates that this is the Green's function for the time-reversed system, where any time-reversal symmetry breaking bias is reversed in direction. A generalized form of reciprocity condition can be written for periodically time-modulated systems \cite{asadchy2020tutorial} (see Supplementary Materials):

\begin{equation}
\frac{\mb{G}^\T_{mp}(\mb{r},\omega_m,\mb{r}',\omega_p)}{\omega_m} = \frac{\mb{G}_{pm}^{\mathcal{T}}(\mb{r}',\omega_p,\mb{r},\omega_m)}{\omega_p}
\label{reciprocityG}
\end{equation}
Using \ref{reciprocityG} into the expression for the partial emissivity we arrive at a relation identical to Eq.~\ref{eq:genKlaw}. In Supplementary Section II we provide numerical verification that this equality holds.

The planar structure results in Fig.~\ref{fig5} were obtained with a one-dimensional time-domain solver for normally incident fields and an effective-medium reduction for oblique incidence. The sawtooth modulation and multilayer geometry are detailed in Supplementary Materials Section III. The absorptivity was calculated from the joule heating and the incident power, and the emissivity was obtained from Eq.~\ref{eq:genKlaw} by reversing the temporal modulation.


\onecolumngrid
\clearpage

\setcounter{section}{0}
\setcounter{subsection}{0}
\setcounter{subsubsection}{0}
\setcounter{equation}{0}
\setcounter{figure}{0}
\setcounter{table}{0}
\setcounter{secnumdepth}{3}
\renewcommand{\thesection}{\Roman{section}}
\renewcommand{\thesubsection}{\Alph{subsection}}
\renewcommand{\thesubsubsection}{\arabic{subsubsection}}
\renewcommand{\theequation}{S\arabic{equation}}
\renewcommand{\thefigure}{S\arabic{figure}}
\renewcommand{\thetable}{S\arabic{table}}
\renewcommand{\theHsection}{supplement.\Roman{section}}
\renewcommand{\theHsubsection}{supplement.\Roman{section}.\Alph{subsection}}
\renewcommand{\theHsubsubsection}{supplement.\Roman{section}.\Alph{subsection}.\arabic{subsubsection}}
\renewcommand{\theHequation}{supplement.\arabic{equation}}
\renewcommand{\theHfigure}{supplement.\arabic{figure}}
\renewcommand{\theHtable}{supplement.\arabic{table}}
\renewcommand{\figurename}{Supplementary Figure}

\begin{center}
{\supplementtitleformat
Supplementary Materials for\\[0.7em]
A generalized Kirchhoff's law of thermal radiation for Floquet media\par}
\vspace{1.2em}
{\supplementauthorformat
Sander A. Mann,\textsuperscript{1,2}
Dimitrios L. Sounas,\textsuperscript{3}
and Andrea Al\`u\textsuperscript{2,4}\par}
\vspace{0.9em}
{\supplementaffiliationformat
\textsuperscript{1}Institute of Physics, University of Amsterdam, Amsterdam, The Netherlands\par
\textsuperscript{2}Photonics Initiative, Advanced Science Research Center, City University of New York, New York, NY 10031, USA\par
\textsuperscript{3}Department of Electrical and Computer Engineering, Wayne State University, Detroit, MI, USA\par
\textsuperscript{4}Physics Program, Graduate Center, City University of New York, New York, NY 10016, USA\par}
\end{center}

\section*{Contents}

\begingroup
\setlength{\parindent}{0pt}
\newcommand{\supptocsection}[3]{%
  \textbf{#1\quad #2}\dotfill\pageref{#3}\par}
\newcommand{\supptocsubsection}[3]{%
  \hspace*{2em}#1\quad #2\dotfill\pageref{#3}\par}
\newcommand{\supptocsubsubsection}[3]{%
  \hspace*{4em}#1\quad #2\dotfill\pageref{#3}\par}
\supptocsection{I}{Coupled-mode theory derivation of the generalized Kirchhoff's law}{sm:sec:cmt}
\supptocsubsection{A}{The Floquet system}{sm:subsec:floquet}
\supptocsubsection{B}{Floquet symmetries and identities}{sm:subsec:symmetries}
\supptocsubsection{C}{Absorptivity and emissivity}{sm:subsec:ae}
\supptocsubsubsection{1}{Absorptivity matrix}{sm:subsubsec:abs}
\supptocsubsubsection{2}{Emissivity matrix}{sm:subsubsec:emis}
\supptocsubsection{D}{Generalized Kirchhoff's law}{sm:subsec:gkl}
\supptocsection{II}{Full-wave derivation}{sm:sec:fullwave}
\supptocsubsection{A}{Cross-correlation operators}{sm:subsec:cross}
\supptocsubsection{B}{Numerical verification}{sm:subsec:num}
\supptocsection{III}{Planar structure demonstrating near-complete violation of Kirchhoff's law}{sm:sec:planar}
\supptocsection{}{Supplemental References}{sm:references}
\endgroup

\section{Coupled-mode theory derivation of the generalized Kirchhoff's law}\label{symmetrysection}\label{sm:sec:cmt}\noindent
We write the equation of motion for $k$ coupled resonators coupled to $l$ external ports and $n$ dissipative loss channels as

\begin{equation}
\frac{d}{dt} \mb{a} = -i\bigl(
\mb{H}(t) - i  \mb{\Gamma} \bigr) \mb{a}  +
\mb{K} ^\T \mb{n}_{e}^+   +  
\mb{L} ^\T \mb{n}_{i} 
\label{general_eom}
\end{equation}
Here $\mb{a}$ is a complex mode amplitude, $\mb{H}(t)$ is a time-varying $k \times k$ matrix, $\mb{\Gamma}$ is time-invariant and also $k\times k$. The interaction with external ports is given by the $l \times k$ matrix $\mb{K}$, while the excitation due to absorption loss-related fluctuations are governed by the $n \times k$ matrix $\mb{L}$, with $\mb{n}_e^+$ incoming thermal radiation and $\mb{n}_i$ is a stochastic source due to fluctuating currents inside the resonators.

The loss matrices are given by

\begin{gather}
\mb{\Gamma}=\mb{\Gamma}_i + \mb{\Gamma}_e \\
\mb{L} ^\dagger \mb{L} = 2\mb{\Gamma}_{i} \\
\mb{K} ^\dagger \mb{K} = 2\mb{\Gamma}_{e}
\end{gather}
The stochastic output is given by

\begin{equation}
\mb{n}_{e}^- = \mb{C} \mb{n}_{e}^+ + \mb{D} \mb{a}.
\end{equation}
These matrices adhere to the following identities:

\begin{equation}
\mb{C}^\T\mb{D}^* = -\mb{K}, \qquad
\mb{D} = \mb{K}^\mathcal{T}, \qquad
\mb{K} = \mb{D}^\mathcal{T}, \qquad
\mb{C}^\mathcal{T} = \mb{C}^\T, \qquad
\mb{\Gamma}^\dagger = \mb{\Gamma}
\end{equation}
where the superscript $\mathcal{T}$ refers to the adjoint system which has time-reversed bias (e.g. reversed magnetic bias $B \rightarrow -B)$. In reciprocal systems with a time-invariant modal basis, some of these identities further simplify to

\begin{equation}
\mb{C}\mb{D}^* = -\mb{K}, \qquad 
\mb{D} = \mb{K}, \qquad
\mb{C} = \mb{C}^\T.
\end{equation}

\subsection{The Floquet system}\label{sm:subsec:floquet}\noindent
For a system where $\mb{H}$ is time-modulated with period $T=\frac{2\pi}{\Omega}$, we expand $\mb{H}$ as

\begin{equation}
\mb{H}(t) = \sum \mb{H}_n e^{-in\omega_mt},
\end{equation}
with $\mb{H}_n = \frac{1}{T} \int_0^T \mb{H}(t) e^{in\omega_mt}dt$.  For the Floquet system, we will consider a stacking of harmonics as $$\hat{\mb{a}} = (\cdots,\mb{a}_{+,1},\mb{a}_{+,0},\mb{a}_{+,-1},\cdots)^\T,$$ where the hat refers to the Floquet basis. In this basis, $\mbh{a}$ is normalized such that $\mbh{a}^\dagger \mbh{a}$ gives the total number of photons in the system and $\hbar \mbh{a} \mbh{\Omega} \mbh{a}$ is the total energy, where $\mbh{\Omega} = \text{diag}(\omega_n \mb{1}_{k\times k} )$ is a block diagonal matrix containing the frequencies of the harmonics, $\omega_n = \omega + n \Omega$. The motivation to work in this basis is that $\mbh{H}$ is Hermitian when parametric gain processes can be ignored. This yields the equation of motion

\begin{equation}
-i \mbh{\Omega} \hat{\mb{a}}  = - (i \hat{\mb{H}} + \hat{\mb{\Gamma}})\hat{\mb{a}} + \mbh{K} ^\T \mbh{n}_{e,+}   +  
\mbh{L} ^\T \mbh{n}_{i,+} ,
\end{equation}
where the input vectors are given by $$\mbh{n}_{j} = (\cdots,\mb{n}_{j,1},\mb{n}_{j,0},\mb{n}_{j,-1},\cdots)^\T,$$ with $\langle \mb{n}_{i,k} (\omega) \mb{n}_{j,l} (\omega') \rangle =\frac{1}{2\pi}  (\Theta(\omega_k,T)/\hbar \omega) \delta_{ij}\delta_{kl} \delta(\omega-\omega')$, where $\Theta(\omega,T) = \hbar \omega /(\exp(\hbar \omega/k_B T)-1)$ is the mean energy of the classical oscillator and is evaluated at frequency $\omega_k$ of the $k$th harmonic. For $\mbh{H}$, we have

\begin{equation}
\mbh{H} = [\mb{H}_m]_T 
\end{equation}
where the subscript $T$ refers to Toeplitz block matrix form, i.e. for a matrix $\mb{M}_m$ indexed by harmonic $m$ we have $([\mb{M}_m]_T)_{k,l} = \mb{M}_{k-l}$. Since the other matrices are not time-modulated, we simply have

\begin{equation}
\mbh{\Gamma} = [\mb{\Gamma}]_D, \qquad
\mbh{K}_e = [\mb{K}_e]_D, \qquad
\mbh{K}_i = [\mb{K}_i]_D, \qquad
\mbh{C} = [\mb{C}]_D,  
\end{equation}
where $[\mb{M}]_D = \mathds{1}\otimes \mb{M} $, i.e. the same as the Toeplitz form but with only the fundamental non-zero.  We can readily solve for the Floquet mode amplitudes

\begin{equation}
\hat{\mb{a}}  = ( i (\mbh{H} -  \mbh{\Omega}) - \mb{\hat{\Gamma}})^{-1} ( \mbh{K} ^\T \mbh{n}_{e}^+   +  \mbh{L}^\T \mbh{n}_{i})
\end{equation}
This yields for reflected wave

\begin{equation}
\mbh{n}_e^- = \mbh{C}\mbh{n}_e^+ + \mbh{D} \mbh{a}.
\end{equation}
Assuming for a moment that internal processes are negligible ($\mbh{n}_i = 0$) enables us to identify the scattering matrix $\mbh{n}_e^- = \mbh{S} \mbh{n}_e^+$ as

\begin{equation}
\mbh{S} = \mbh{C} + \mbh{D}  ( i (\mbh{H} -  \mbh{\Omega}) - \mb{\hat{\Gamma}})^{-1} \mb{\hat{K}^\T} =  \mb{\hat{C}} + \mb{\hat{D}} \mb{\hat{G}} \mb{\hat{K}^\T} 
\end{equation}
where we have defined the Floquet Green's function $\mbh{G}$ as

\begin{equation}
\mbh{G}  = ( i (\hat{\mb{H}} -  \hat{\mb{\Omega}}) - \mb{\hat{\Gamma}})^{-1}. 
\end{equation}

\subsection{Floquet symmetries and identities}\label{sm:subsec:symmetries}\noindent
Since most coefficient matrices are block diagonal, the same identities hold as before:

\begin{gather}
\mbh{C}^\T\mbh{D}^* = -\mbh{K}, \qquad
\mbh{D} = \mbh{K}^\mathcal{T}, \qquad
\mbh{K} = \mbh{D}^\mathcal{T}, \qquad
\mbh{C}^\mathcal{T} = \mbh{C}^\T, \qquad
\mbh{\Gamma}^\dagger = \mbh{\Gamma}.
\end{gather}
For $\mbh{H}$ we, have $\mb{H}_n^\dagger = \mb{H}_{-n}$ which follows from the expansion for $\mb{H}(t)$, which indeed yields $\mbh{H}^\dagger = \mbh{H}$. It is also useful to consider the impact of the time-reversal symmetry breaking biases. We may distinguish between two types: static biases (such as a magnetic field) and time-reversal symmetry breaking due to temporal modulation. In the following, we will denote static bias with $\lambda$ and reversed temporal modulation with $-t$, i.e. the conservative generator with both biases reversed would read $\mb{H}(-t,-\lambda)$. The impact of the reversal of the static bias can be understood by considering the Onsager-Casimir reciprocal relation at each instant $t$, i.e. $$\mb{H}(t,-\lambda)= \mb{H}^\T(t,\lambda).$$ The impact of reversed modulation (superscript r) can be understood from the expansion $$\mb{H}^\text{r} = \mb{H}(-t) = \sum \mb{H}_n e^{-in\omega_m t} = \sum \mb{H}_{-n} e^{in\omega_m t},$$ which means $$\mb{H}^\text{r}_n = \mb{H}_{-n}.$$ We will be interested in the relationship between a system and its adjoint counterpart, a system identical but for its biases which are all reversed. We will denote such adjoint systems by the superscript $\mathcal{T}$. Based on the aforementioned, for the microreversed system where both both the direction of the static bias and the temporal modulation are reversed, we obtain 

\begin{equation}
\mbh{H}^\mathcal{T} = \mbh{H}^\T.
\end{equation}
While we don't consider temporal modulation in the losses, Onsager-Casimir reciprocity still applies, which yields $$\mbh{\Gamma}(-\lambda) = \mbh{\Gamma}^\T(\lambda)$$.

\subsection{Absorptivity and emissivity}\label{sm:subsec:ae}\noindent
To find the absorptivity and emissivity matrices we consider the system with zero internal and external temperature, respectively, such that each process can be considered in isolation. We are after a relationship between the absorptivity and emissivity matrices, an extension of the time-invariant Kirchhoff's law. In the context of time-varying systems with frequency conversion, we explicitly define the absorbed power as power incident at $\omega$ that is dissipated in absence of excitation from internal sources. This yields a definition for the absorptivity matrix $\boldsymbol{\alpha}$ as a unitless matrix that gives the absorbed power when left- and right-multiplied by input vectors, \emph{i.e.} $P_\text{abs} = \mb{s}^\dagger \boldsymbol{\alpha} \mb{s}$ where $\mbh{s}$ is an input vector normalized such that $\mbh{s}^\dagger \mbh{s}$ is power. Likewise, for the emissivity matrix, we consider $\boldsymbol{\epsilon}$ a unitless matrix that gives the emitted power at frequency $\omega$ when left- and right multiplied by an output vector $P_\text{em} = \mb{s}^\dagger \boldsymbol{\epsilon} \mb{s}$.

\subsubsection{Absorptivity matrix}\label{sm:subsubsec:abs}\noindent
In analogy with the standard expression $P_{abs} = 2\mb{a}^\dagger \mb{\Gamma} \mb{a}$ in time-invariant coupled-mode theory (in the energy basis), we can write in the Floquet system 

\begin{equation}
P_\text{abs}(\omega) = \hbar  \mbh{a}^\dagger \mbh{L}^\dagger \mbh{\Omega} \mbh{L} \mbh{a}.
\end{equation}
Inserting the expression for the cavity amplitude (in absence of $\mbh{n}_i$), we find

\begin{equation}
P_\text{abs}(\omega) = \hbar \langle \mb{n}_e^\dagger \mbh{K}^\ast \mbh{G}^\dagger \mbh{L}^\dagger \mbh{\Omega} \mbh{L} \mbh{G} \mbh{K}^\T \mbh{n}_e \rangle
\end{equation}
Armed with this expression, we can work towards the absorptivity matrix. First, let us define the Floquet absorptivity matrix that captures all harmonics:

\begin{equation}
\hat{\boldsymbol{\alpha}} = \mbh{\Omega}^{-\frac{1}{2}} \mbh{K}^* \mbh{G}^\dagger \mbh{L}^\dagger \mbh{\Omega} \mbh{L} \mbh{G} \mbh{K}^\T \mbh{\Omega}^{-\frac{1}{2}}.
\end{equation}
where $\hat{\boldsymbol{\alpha}}_{lm}$ is the absorptivity matrix due to excitation at the $l$-th and $m$-th harmonic simultaneously. We multiply by $(\hbar \mbh{\Omega})^{-\frac{1}{2}}$ on the left and the right because as we are interested in an absorptivity matrix normalized for \emph{power}, while our Floquet formalism is normalized for quanta. The absorptivity matrix $\boldsymbol{\alpha}$ due to excitation at the fundamental frequency is given by $\boldsymbol{\alpha} = \hat{\boldsymbol{\alpha}}_{00}$, which we can write as

\begin{equation}
\boldsymbol{\alpha} = \hat{\boldsymbol{\alpha}}_{00} = \sum_m (\mbh{\Omega}^{-\frac{1}{2}} \mbh{K}^* \mbh{G}^\dagger \mbh{L}^\dagger)_{0m} (\mbh{\Omega} \mbh{L} \mbh{G} \mbh{K}^\T \mbh{\Omega}^{-\frac{1}{2}})_{m0} =  \sum_m   \hat{\boldsymbol{\alpha}}_m,
\end{equation}
where we have defined 

\begin{equation}
\hat{\boldsymbol{\alpha}}_m = \frac{\omega_m}{\omega}  (\mbh{K}^* \mbh{G}^\dagger \mbh{L}^\dagger)_{0m} ( \mbh{L} \mbh{G} \mbh{K}^\T )_{m0},
\label{eq:partialabsorptivity}
\end{equation}
and we have made use of the fact that $\mbh{\Omega}$ is block diagonal, which allows us to pull it out of the matrix product into the sum. We can identify the sum over submatrices here as the partial absorptivity in different harmonics $\hat{\boldsymbol{\alpha}}_m$ due to excitation at the fundamental frequency. The ratio $\frac{\omega_m}{\omega}$ captures the increase in absorbed power if incident quanta are first upconverted, and vice versa.

\subsubsection{Emissivity matrix}\label{sm:subsubsec:emis}\noindent
We define as the emission at frequency $\omega$ into the normalized vector $\mb{u}$ due to anything that occurs within the system \emph{in absence of incident thermal radiation} as:

\begin{equation}
P_\text{em} (\omega) = \hbar \omega \mb{u}^\dagger \langle \mbh{n}_{e}^-  (\mbh{n}_{e}^-)^\dagger \rangle _{00}\mb{u}
\end{equation}
Writing out the ensemble average using $\mbh{n}_{e}^- = \mbh{D} \hat{\boldsymbol{\alpha}} = \mbh{D} \mbh{G} \mbh{L}^\T \mbh{n}_i$ gives

\begin{equation}
 \langle \mbh{n}_{e}^-  (\mbh{n}_{e}^-)^\dagger \rangle _{00} = \sum_m (\mbh{D} \mbh{G} \mbh{L}^\T)_{0m}  \langle \mbh{n}_i \mbh{n}_i^\dagger \rangle_{mm} (\mbh{L}^* \mbh{G}^\dagger \mbh{D}^\dagger )_{m0}.
\end{equation}
Using $ \langle \mbh{n}_i \mbh{n}_i^\dagger \rangle_{pq} =\frac{1}{2\pi} (\Theta(\omega_p,T)/\hbar \omega_p) \delta_{pq}$ we can define the emitted power due to fluctuations at harmonic $m$

\begin{equation}
P_\text{em,m} (\omega,T) = \frac{\Theta(\omega_m,T)}{2\pi}  \mb{u}^\dagger \frac{\omega}{\omega_m} (\mbh{D} \mbh{G} \mbh{L}^\T)_{0m} (\mbh{L}^* \mbh{G}^\dagger \mbh{D}^\dagger )_{m0} \mb{u}
\end{equation}
Here, we can identify the power-basis partial emissivity as

\begin{equation}
\hat{\boldsymbol{\epsilon}}_m = \frac{\omega}{\omega_m} (\mbh{D} \mbh{G} \mbh{L}^\T)_{0m} (\mbh{L}^* \mbh{G}^\dagger \mbh{D}^\dagger )_{m0}
\end{equation}
in analogy with the partial absorptivity in Eq.~\eqref{eq:partialabsorptivity}. To find the total emissivity, we take the total emitted power and normalize by Planck's law at the fundamental, which yields for the emissivity matrix for emission at $\omega$:

\begin{equation}
\boldsymbol{\epsilon} =  \sum_m \frac{ \Theta(\omega_m,T)}{ \Theta(\omega,T)} \hat{\boldsymbol{\epsilon}}_m.
\end{equation}

\subsection{Generalized Kirchhoff's law}\label{sm:subsec:gkl}\noindent
In a previous section we have shown that under bias reversal, $\mbh{\Gamma} \rightarrow \mbh{\Gamma}^\T $ and $\mbh{H} \rightarrow \mbh{H}^\T$. This means that $\mbh{G} \rightarrow \mbh{G}^\T$ under bias reversal, given that the inverse of a transpose is the transpose of an inverse. Applying reversal of the bias to the harmonic absorptivity, we thus find

\begin{equation}
 \frac{\omega_m}{\omega}  (\mbh{K}^* \mbh{G}^\dagger \mbh{L}^\dagger)_{0m} (\mbh{L} \mbh{G} \mbh{K}^\T)_{m0} \rightarrow  \frac{\omega_m}{\omega}  (\mbh{D}^* \mbh{G}^* \mbh{L}^\dagger)_{0m} (\mbh{L} \mbh{G}^\T \mbh{D}^\T)_{m0}
\end{equation}
which we will denote as $\hat{\boldsymbol{\alpha}}_m^\mathcal{T}$. We can recognize the matrix product on the right hand side as the matrix product in $\hat{\boldsymbol{\epsilon}}_m^*$, which means we find a time-reversal symmetry relationship between the partial absorptivities and emissivities: $\hat{\boldsymbol{\epsilon}}_m =  \frac{\omega^2}{\omega_m^2} (\hat{\boldsymbol{\alpha}}_m^{\mathcal{T}})^*$.  This leads us to the generalized Kirchhoff's law:

\begin{equation}
\boldsymbol{\epsilon} = \sum_m \frac{\Theta(\omega_m,T)}{\Theta(\omega,T)} \frac{\omega^2}{\omega_m^2}  \bigl(\boldsymbol{\alpha}_m^{\mathcal{T}}\bigr)^*
\end{equation}
In absence of modulation, only the fundamental $m=0$ persists, which results in the standard Kirchhoff's law.

\section{Full-wave derivation}\label{sm:sec:fullwave}
\noindent
Thermal radiation is the result of fluctuation currents in lossy media. In time-invariant media under thermal equilibrium, the cross-correlation function of fluctuation currents is given by the fluctuation-dissipation theorem (FDT),
\begin{equation}
\label{eq:FDT}
    \big\langle \Jbf(\rbf,\omega) \Jbf^\dagger(\rbf',\omega') \big\rangle = 4\pi \omega \Theta(|\omega|) \bm{\varepsilon}_\mathrm{I}(\rbf,\omega)  \delta(\rbf-\rbf') \delta(\omega - \omega'),
\end{equation}
where $\bm{\varepsilon}_\mathrm{I}(\rbf,\omega) = [\bm{\varepsilon}(\rbf,\omega) - \bm{\varepsilon}^\dagger(\rbf,\omega)]/(2\im)$ is the Hermitian imaginary part of $\bm{\varepsilon}(\rbf,\omega)$, $\Theta(\omega) = \hbar\omega / [ \exp(\hbar\omega/(kT)) - 1]$ is the mean energy of the classical oscillator, $\hbar$ is the reduced Planck's constant, $k$ the Boltzmann constant, and $T$ the temperature \cite{supp-landau-lifshitz-course}. Time-modulated media are generally not in local equilibrium and therefore Eq.~\eqref{eq:FDT} does not  apply to them. However, it still applies to dissipative unmodulated media embedded in time-modulated media. Here, we focus on this case and analyze thermal emission from heterogeneous structures consisting of time-invariant lossy media and time-varying lossless media. In this case, thermal emission is generated by the time-invariant parts of the structure. The more general case of thermal radiation from the time-varying regions of the structure requires a revision of the FDT and is beyond the scope of this paper.


To find the statistical properties of the radiated field we need to calculate the cross-correlation function $\langle \Ebf(\rbf,\omega) \Ebf^\dagger(\rbf',\omega')\rangle$. Replacing the electric field with its Green function integral we find
\begin{equation}
    \blangle \Ebf(\rbf,\omega) \Ebf^\dagger (\rbf',\omega') \brangle = 4\pi \intt \intV \omega'' \Theta(|\omega''|) \mb{G}(\rbf,\omega;\rbf'',\omega'') \bm{\varepsilon}_\mathrm{I}(\rbf'',\omega'') \mb{G}^\dagger (\rbf',\omega';\rbf'',\omega'') \,\dV'' \domega'', \label{eq:x-correlation-TMM}
\end{equation}
where $\mb{G}(\rbf,\omega;\rbf',\omega')$ is the Green function that converts a current at point $\rbf'$ and frequency $\omega'$ to an electric field at point $\rbf$ and frequency $\omega$. For periodically time-modulated media (PTMM), fields at frequency $\omega$ are generated by currents at frequencies $\omega_n = \omega + n\Omega$, where $\Omega$ is the modulation frequency, and the Green's function becomes
\begin{equation}
    \mb{G}(\rbf,\omega;\rbf',\omega') = \sum_n \mb{G}_{n}(\rbf,\rbf',\omega) \delta(\omega'-\omega_n), \label{eq:FDGF-PTMM}
\end{equation}
where $\mb{G}_{n}(\rbf,\rbf',\omega)$ is the Green's function that describes conversion from currents at frequency $\omega_n$ to fields at frequency $\omega$. Replacing Eq.~\eqref{eq:FDGF-PTMM} into Eq.~\eqref{eq:x-correlation-TMM} gives
\begin{equation}
    \blangle \Ebf(\rbf,\omega)\Ebf^\dagger (\rbf',\omega') \brangle = 4\pi \sum_{mn} \delta(\omega'-\omega_m) \omega_n \Theta(|\omega_n|) \intV \mb{G}_{n}(\rbf,\rbf'',\omega) \bm{\varepsilon}_\mathrm{I}(\rbf'',\omega_n)  \mb{G}^\dagger_{n-m} (\rbf',\rbf'',\omega_m) \,\dV''. \label{eq:x-correlation-PTMM}
\end{equation}
This equation shows that the radiated fields can be correlated if they have frequencies that are different by an integer multiple of the modulation frequency. This happens because such fields are generated by the same fluctuation currents. 

Eq.~\eqref{eq:x-correlation-PTMM} is the cross-correlation function of a cyclo-stationary process. If $x(t)$ is such a process with period $T$, the two-frequency spectral correlation function $\langle X(\omega) X^\ast(\omega') \rangle$ is a sum of delta functions $\delta(\omega'-\omega_n)$ as
\begin{equation}
    \langle X(\omega) X^\ast(\omega') \rangle = \sum_n S_n(\omega) \delta(\omega'-\omega_n), \label{eq:x-corr-general-process}
\end{equation}
where $S_n(\omega)$ are smooth functions of $\omega$. The ensemble averaged power is
\begin{equation}
    \langle|x(t)|^2\rangle = \frac{1}{4\pi^2} \sum_n e^{-\im n\Omega t} \int_{-\infty}^\infty S_n(\omega) \domega.
\end{equation}
This equation shows that the power is periodic in time with harmonic components
\begin{equation}
    p_n = \frac{1}{4\pi^2} \int_{-\infty}^\infty S_n(\omega) \domega.
\end{equation}
The time-averaged power is found from $n=0$ and has a two-sided power spectral density
\begin{equation}
    W_0(\omega) = \frac{S_0(\omega)}{4\pi^2}.
\end{equation}
Similarly, $W_n(\omega) = S_n(\omega)/(4\pi^2)$ is the power spectral density for the $n$-th harmonic component of the ensemble averaged power.

Applying this result to Eq.~\eqref{eq:x-correlation-PTMM}, we find the cross-spectral density tensor at point $\rbf$ and frequency $\omega$ as
\begin{equation}
    \mb{W}(\rbf,\omega) = \frac{1}{\pi} \sum_n \intV \omega_n \Theta(|\omega_n|) \mb{G}_{n}(\rbf,\rbf',\omega)
    \bm{\varepsilon}_\mathrm{I}(\rbf',\omega_n) \mb{G}_{n}^\dagger (\rbf,\rbf',\omega) \,\dV'. \label{eq:xSD-PTMM}
\end{equation}
To find the spectral density for polarization $\uvec{p}$ we project $\mb{W} (\rbf,\omega)$ on $\uvec{p}$ resulting in the polarization-resolved spectral density
\begin{equation}
    \label{eq:p-SD-PTMM}
    W_\mb{p}(\rbf,\omega) = \uvec{p}^\dagger \mb{W}(\rbf,\omega) \uvec{p}.
\end{equation}
At large distances from the object, the radiated field takes the form of a plane wave, and the radiated power per unit solid angle in direction $\uvec{r}$ and for polarization $\uvec{p}$ is
\begin{equation}
    \label{eq:Wrad-PTMM}
    W_\mathrm{rad}(\uvec{r},\hat{\mathbf{p}},\omega) = \frac{2}{\eta_0} \lim_{r\to\infty} r^2 \uvec{p}^\dagger \mb{W}(\rbf,\omega) \uvec{p},
\end{equation}
where the factor of $2$ accounts for contributions from both sides of the spectrum. Replacing Eq.~\eqref{eq:xSD-PTMM} into this equation gives
\begin{equation}
    W_\mathrm{rad}(\uvec{r},\hat{\mathbf{p}},\omega) = \frac{2}{\pi\eta_0} \lim_{r\to\infty} r^2 \sum_n \intV \omega_n \Theta(|\omega_n|) \uvec{p}^\dagger \mb{G}_{n}(\rbf,\rbf',\omega) \bm{\varepsilon}_\mathrm{I}(\rbf',\omega_n) \mb{G}^\dagger_{n} (\rbf,\rbf',\omega) \uvec{p} \,\dV'. \label{eq:Wrad-PTMM-1}
\end{equation}

To find the emissivity, we divide the radiated power by the polarization resolved black-body spectral radiance 
\begin{equation}
    B_\mathrm{p}(\omega) = \frac{\omega^2}{8\pi^3 c^2} \Theta(\omega).
\end{equation}
This yields
\begin{equation}
    \label{eq:emissivity-PTMM}
    \epsilon(\uvec{r},\hat{\mathbf{p}},\omega) = \sum_n \frac{\Theta(|\omega_n|)}{\Theta(\omega)} \epsilon_n(\uvec{r},\hat{\mathbf{p}},\omega)
\end{equation}
where
\begin{equation}
    \epsilon_n(\uvec{r},\hat{\mathbf{p}},\omega) = \frac{16\pi^2 c}{\omega\mu_0} \lim_{r\to\infty} r^2 \intV \frac{\omega_n}{\omega} \uvec{p}^\dagger \mb{G}_{n}(\rbf,\rbf',\omega) \bm{\varepsilon}_\mathrm{I}(\rbf',\omega_n) \mb{G}^\dagger_{n}(\rbf,\rbf',\omega) \uvec{p} \,\dV' \label{eq:emissivity-PTMM-n}
\end{equation}
is the partial emissivity due to the $n$-order frequency harmonic of the fluctuation currents (i.e., the currents with frequency $\omega_n$).

Now, consider that the object is excited by a point source with frequency $\omega$, polarization $\hat{\mathbf{p}}$, and unit current, at location $\rbf$. The absorbed power is given by the sum of the absorbed power in each of the frequency harmonics as
\begin{equation}
    P_\mathrm{abs} = \frac{1}{2} \sum_n \intV \omega_n \Ebf^\dagger(\rbf',\omega_n) \bm{\varepsilon}_\mathrm{I}(\rbf',\omega_n) \Ebf(\rbf',\omega_n) \,\dV'. \label{eq:Pabs-PTMM}
\end{equation}
From the Green's function in PTMM we know that $\Ebf(\rbf',\omega_n) = \mb{G}_{-n}(\rbf',\rbf,\omega_n) \uvec{p}$. Replacing this expression into Eq.~\eqref{eq:Pabs-PTMM} gives
\begin{equation}
    P_\mathrm{abs} = \frac{1}{2} \sum_n \intV \omega_n \uvec{p}^\dagger \mb{G}_{-n}^\dagger(\rbf',\rbf,\omega_n) \bm{\varepsilon}_\mathrm{I}(\rbf',\omega_n) \mb{G}_{-n}(\rbf',\rbf,\omega_n)\uvec{p} \,\dV'. \label{eq:Pabs-PTMM-1}
\end{equation}
Assuming that the source is at a large distance from the object, the incident electric field on the object can be approximated as a plane wave with an amplitude
\begin{equation}
    E_\mathrm{inc} = \frac{\im \omega \mu_0}{4\pi} \frac{e^{-\im kr}}{r}.
\end{equation}
The power density of this wave is
\begin{equation}
    P_\mathrm{inc} = \frac{\omega^2\mu_0}{32\pi^2cr^2}.
\end{equation}
Dividing $P_\mathrm{abs}$ by $P_\mathrm{inc}$ yields the absorption cross section for excitation from direction $-\uvec{r}$ and polarization $\uvec{p}$,
\begin{equation}
    \alpha(-\uvec{r},\uvec{p},\omega) = \sum_n \alpha_n(-\uvec{r},\uvec{p},\omega),
\end{equation}
where
\begin{equation}
    \alpha_n(-\uvec{r},\uvec{p},\omega) = \frac{16\pi^2c}{\omega\mu_0} \lim_{r\to\infty} r^2 \intV \frac{\omega_n}{\omega}    \uvec{p}^\dagger \mb{G}_{-n}^\dagger(\rbf',\rbf,\omega_n) \bm{\varepsilon}_\mathrm{I}(\rbf',\omega_n)  \mb{G}_{-n}(\rbf',\rbf,\omega_n) \uvec{p} \,\dV' \label{eq:absorptivity-PTMM}
\end{equation}
is the partial absorption cross section through the $n$-order harmonic of the induced currents.

Although PTMM generally break Lorentz reciprocity, they satisfy the generalized reciprocity theorem
\begin{equation}
    \label{eq:reciprocity-PTMM-2}
    \sum_n \intV \frac{\Jbf(\rbf,\omega_n) \cdot \Ebf^{\mathcal{T}}(\rbf,\omega_n)}{\omega_n} \, \dV = \sum_n \intV \frac{\Jbf^{\mathcal{T}}(\rbf,\omega_n) \cdot \Ebf(\rbf,\omega_n)}{\omega_n} \, \dV,
\end{equation}
where $\Jbf(\rbf,\omega_n)$ and $\Ebf(\rbf,\omega_n)$ are the electric currents and fields of the original system, and $\Jbf^{\mathcal{T}}(\rbf,\omega_n)$ and $\Ebf^{\mathcal{T}}(\rbf,\omega_n)$ are the electric currents and fields of the time-reversed system. Eq.~\eqref{eq:reciprocity-PTMM-2} was derived in \cite{supp-asadchy2020tutorial} for lossless media, but it can be shown to be valid in lossy media with time invariant loss by adding the loss currents to $\Jbf(\rbf,\omega_n)$ and $\Jbf^{\mathcal{T}}(\rbf,\omega_n)$, and showing that the loss terms cancel out. Here, $\mathcal{T}$ refers to a restricted time reversal that reverses time and odd-parity biases, but it does not swap loss with gain. Replacing 
\begin{equation}
    \Ebf(\rbf,\omega_n) = \sum_m \int_V \mb{G}_{m-n}(\rbf,\rbf',\omega_n) \Jbf(\rbf,\omega_m)\dV'
\end{equation}
into Eq.~\eqref{eq:reciprocity-PTMM-2}, we find a reciprocity relation for Green's functions as
\begin{equation}
    \frac{\mb{G}_{n-m}^\mathcal{T}(\rbf,\rbf',\omega_m)}{\omega_m} = \frac{\mb{G}_{m-n}^\T(\rbf',\rbf,\omega_n)}{\omega_n}. \label{eq:reciprocity-PTMM-3}
\end{equation}
Replacing $m=0$ in this equation yields
\begin{equation}
    \mb{G}_{n}(\rbf,\rbf',\omega) = \frac{\omega}{\omega_n} \mb{G}_{-n}^{\mathcal{T}\T}(\rbf',\rbf,\omega_n).
\end{equation}
Replacing this equation into Eq.~\eqref{eq:emissivity-PTMM-n} results in
\begin{equation}
    \epsilon_n(\uvec{r},\hat{\mathbf{p}},\omega) = \frac{16\pi^2 c}{\omega\mu_0}
    \lim_{r\to\infty} r^2 \intV \frac{\omega}{\omega_n} \uvec{p}^\dagger \mb{G}_{-n}^{\mathcal{T}\T}(\rbf',\rbf,\omega_n) \bm{\varepsilon}_\mathrm{I}(\rbf',\omega_n) \mb{G}^{\mathcal{T}\ast}_{-n}(\rbf',\rbf,\omega_n) \uvec{p} \,\dV'. \label{eq:emissivity-PTMM-n-1}
\end{equation}
Comparing this equation with Eq.~\eqref{eq:absorptivity-PTMM} we find
\begin{equation}
    \epsilon_n(\uvec{r},\hat{\mathbf{p}},\omega) = \frac{\omega^2}{\omega_n^2} \alpha_n^\mathcal{T}(-\uvec{r},\uvec{p}^\ast,\omega). \label{eq:Generalized-Kirchhoff}
\end{equation}
This equation shows that the $n$-order partial emissivity of the original system in direction $\uvec{r}$, polarization $\uvec{p}$ and frequency $\omega$ is equal to the $n$-order absorptivity of the time-reversed system in the opposite direction $-\uvec{r}$, conjugate polarization $\uvec{p}^\ast$, and the same frequency $\omega$, scaled by the factor $\omega^2/\omega_n^2$. This equation is the generalization of Kirchhoff's law of thermal radiation to PTMM. Similarly to nonreciprocal time-invariant media \cite{supp-guo2022adjoint}, Kirchhoff's law in PTMM applies between a medium and the time reversed medium. However, in PTMM Kirchhoff's law holds separately for the partial emissivities and absorptivities at frequency harmonics instead of for the total emissivity and absorptivity as in time-invariant media.

Eq.~\eqref{eq:Generalized-Kirchhoff} offers a convenient way to calculate thermal emission in PTMM in cases where the Green function is not analytically known and the analysis needs to be performed numerically. In such cases, the calculation of $\alpha_{n}^\mathcal{T}(-\uvec{r},\uvec{p}^\ast,\omega)$ requires only one simulation for a specific frequency, a specific polarization, and a specific propagation direction, in contrast to $\epsilon_n(\uvec{r},\hat{\mathbf{p}},\omega)$, which requires the calculation of the radiated field from sources at any point within the medium. Therefore, the generalized Kirchhoff law significantly simplifies the calculation of thermal radiation in PTMM.

\subsection{Cross-correlation operators}\label{sm:subsec:cross}
Kirchhoff's law can also be expressed in terms of emissivity and absorptivity operators. In TIM, these operators give correlations of emitted or absorbed fields in different directions. In PTVM, they also give correlations in frequency. 

In TIM, the emissivity operator can be defined as
\begin{equation}
    \boldsymbol{\epsilon}(\uvec{\rbf},\uvec{\rbf}',\omega) = \frac{2}{\eta_0 B_\mathrm{p}(\omega)} \lim_{r,r'\to\infty} rr' \mb{W}(\rbf,\rbf',\omega), \label{eq:emissivity-op-TIM}
\end{equation}
where $\mb{W}(\rbf,\rbf')$ is the power cross spectral density of the radiated fields at $\rbf$ and $\rbf'$, defined as
\begin{equation}
    \mb{W}(\rbf,\rbf',\omega) = \lim_{T_0\to\infty} \frac{T_0}{2\pi} \langle\Ebf_{T_0}(\rbf,\omega)\Ebf_{T_0}^\dagger(\rbf',\omega)\rangle, \label{eq:powe-x-spectral-density-stationary}
\end{equation}
where
\begin{equation}
    \Ebf_{T_0}(\omega) = \frac{1}{T_0} \int_{-T_0/2}^{T_0/2} \Ebf(t) e^{-\im\omega t}\dt
\end{equation}
is the time-averaged complex spectral amplitude at frequency $\omega$. Eq.~\eqref{eq:powe-x-spectral-density-stationary} follows from the fact that in the limit of large $T_0$, $\langle\Ebf_{T_0}(\rbf,\omega)\Ebf_{T_0}^\dagger(\rbf',\omega)\rangle$ is approximately equal to the mutual power in a frequency band of width $\Delta\omega_0 = 2\pi/T_0$ around $\omega$. Dividing this power by $\Delta\omega_0$ yields the power cross spectral density as defined in Eq.~\eqref{eq:powe-x-spectral-density-stationary}. By using the cross-correlation function
\begin{equation}
    \langle\Ebf(\rbf,\omega)\Ebf^\dagger(\rbf',\omega')\rangle = \mb{S}(\rbf,\rbf',\omega)\delta(\omega-\omega') \label{eq:x-correlation-invariant}
\end{equation}
that is valid for any stationary process, it can be shown that
\begin{equation}
    \mb{W}(\rbf,\rbf',\omega) = \frac{\mb{S}(\rbf,\rbf',\omega)}{4\pi^2}.
\end{equation} 
The factor of $2$ in Eq.~\eqref{eq:emissivity-op-TIM} is added to include contributions from both sides of the spectrum. The emissivity operator is Hermitian, and its eigenvectors define orthogonal radiation channels with emissivities equal to the corresponding eigenvalues.

If the object is illuminated by a wave with amplitude $\mathbf{A}(-\uvec{\rbf})$, such that $|\mathbf{A}|^2$ is the power per unit solid angle, the absorptivity operator, denoted by $\boldsymbol{\alpha}(-\uvec{\rbf},-\uvec{\rbf}')$, is defined such that
\begin{equation}
    P_\mathrm{abs} = \int_{4\pi}\int_{4\pi}\mathbf{A}^\dagger(-\uvec{r})\boldsymbol{\alpha}(-\uvec{\rbf},-\uvec{\rbf}')\mathbf{A}(-\uvec{\rbf}')\,\mathrm{d}\Omega\,\mathrm{d}\Omega'
\end{equation}
is the absorbed power. Like the emissivity operator, the absorptivity operator is Hermitian, and its eigenvectors define orthogonal absorption channels with absorptivities equal to the corresponding eigenvalues.

In time invariant media, the emissivity and absorptivity operators are transponse symmetric as
\begin{equation}
    \bm{\epsilon}(\uvec{\rbf},\uvec{\rbf}') = \bm{\alpha}{}^{\mathcal{T}\T}(-\uvec{\rbf}',-\uvec{\rbf}). \label{eq:KLTR-op}
\end{equation}
This result has been derived  through the macroscopic scattering matrix of a scatterer and the principle of detailed balance \cite{supp-miller2017universal}, but it can also be derived from reciprocity and the fluctuation dissipation theorem, as we will show later. A result of Eq.~\eqref{eq:KLTR-op} is that the eigenvectors of $\bm{\epsilon}$ are equal to the complex conjugate eigenvectors of $\bm{\alpha}^\mathcal{T}$. In other words, the radiation channels of an object are equal to the complex conjugate channels of the same object with a reverse odd-symmetric bias. Furthermore, the emissivity in a radiation channel of the original object is equal to the absorptivity of the conjugated absorption channel in the reverse-biased object.

In PTVM, fields belonging to the same frequency ladder are correlated, and the emissivity and absorptivity operators should be defined to reflect this fact. Following Eq.~\eqref{eq:emissivity-op-TIM}, the emissivity operator is defined as
\begin{equation}
    \bm{\epsilon}_{nm}(\uvec{\rbf},\uvec{\rbf}') = \frac{2}{\eta_0} \frac{1}{\sqrt{B_\mathrm{p}(|\omega_n|)B_\mathrm{p}(|\omega_m|)}} \lim_{r,r'\to\infty} rr' e^{\im k_nr} e^{-\im k_mr'} \mb{W}_{nm}(\rbf,\rbf'),
\end{equation}
where $k_n = \omega_n/c$ and $\mb{W}_{nm}(\rbf,\rbf')$ is the power cross-spectral density between the radiated fields at the position-frequency pairs $(\rbf,\omega_n)$ and $(\rbf',\omega_m)$ defined as
\begin{equation}
    \mb{W}_{nm}(\rbf,\rbf') =  \lim_{T_0\to\infty} \frac{T_0}{2\pi} \langle\Ebf_{T_0}(\omega_n)\Ebf_{T_0}^\dagger(\omega_m)\rangle.
\end{equation}
Using the expression
\begin{equation}
    \langle\Ebf(\rbf,\omega)\Ebf^\dagger(\rbf',\omega')\rangle = \sum_n \mb{S}_n(\rbf,\rbf') \delta(\omega'-\omega_n)
\end{equation}
for the cross-correlation function that is valid for any cyclostationary process, it can be shown that
\begin{equation}
    \mb{W}_{nm}(\rbf,\rbf') = \frac{\mb{S}_{m-n}(\omega_n)}{4\pi^2}.
\end{equation}
Applying this result to Eq.~\eqref{eq:x-correlation-PTMM} we find
\begin{equation}
    \mb{W}_{nm}(\rbf,\rbf') = \frac{1}{\pi} \sum_k \omega_k\Theta(|\omega_k|) \int_V \mb{G}_{nk}(\rbf,\rbf'') \bm{\varepsilon}_\mathrm{I}(\rbf'',\omega_k) \mb{G}_{mk}^\dagger(\rbf',\rbf'') \dV'',
\end{equation}
where $\mb{G}_{nk}(\rbf,\rbf') \equiv \mb{G}_{k-n}(\rbf,\rbf',\omega_n)$. Then, 
\begin{equation}
    \bm{\epsilon}_{nm}(\uvec{r},\uvec{r}') = \sum_k \frac{\Theta(|\omega_k|)}{\sqrt{\Theta(|\omega_n|)\Theta(|\omega_m|)}} \bm{\epsilon}_{nmk}(\uvec{r},\uvec{r}'),
\end{equation}
with
\begin{equation}
    \bm{\epsilon}_{nmk}(\uvec{r},\uvec{r}') = \frac{16\pi^2 c}{\mu_0} \frac{\omega_k}{\omega_n\omega_m}  \lim_{r,r'\to\infty} rr' e^{\im k_nr} e^{-\im k_mr'} \int_V \mb{G}_{nk}(\rbf,\rbf'') \bm{\varepsilon}_\mathrm{I}(\rbf'',\omega_k) \mb{G}_{mk}^\dagger(\rbf',\rbf'') \dV''
\end{equation}
the partial absorptivity.

The absorptivity operator is defined such that the absorbed power is
\begin{equation}
    P_\mathrm{abs} = \sum_{nm} \int_{4\pi} \int_{4\pi} \mathbf{A}_n^\dagger(-\uvec{r}) \bm{\alpha}_{nm}(-\uvec{\rbf},-\uvec{\rbf}') \mathbf{A}_m(-\uvec{\rbf}') \,\mathrm{d}\Omega \,\mathrm{d}\Omega', \label{eq:absorbed-power}
\end{equation}
where $\mathbf{A}_n(-\uvec{\rbf})$ is the amplitude of the incident field at frequency $\omega_n$. Assuming that the incident field is generated by dipole sources $\Jbf_n(-\uvec{\rbf}) = \sqrt{2\eta_0} \mathbf{A}_n(-\uvec{\rbf}) 4\pi r e^{\im kr} / (\im\omega_n\mu_0)$ at a large distance $r$ from the object, the electric field at $\omega_n$ is given by
\begin{equation}
    \mathbf{E}_n(\mathbf{r}') = \lim_{r\to\infty} \sum_m \int_{4\pi} \mb{G}_{nm}(\rbf',\rbf) \Jbf_m(-\uvec{\rbf}) \,\mathrm{d}\Omega = \sqrt{2\eta_0} \lim_{r\to\infty} \sum_m \frac{4\pi r}{\im\omega_m\mu_0} e^{\im k_mr} \int_{4\pi} \mb{G}_{nm}(\rbf',\rbf) \mathbf{A}_m(-\uvec{\rbf}) \,\mathrm{d}\Omega.
\end{equation}
By replacing this expression into $P_\mathrm{abs} = \frac{1}{2} \sum_n \int_V \omega_n \Ebf_n^\dagger(\rbf) \bm{\varepsilon}_\mathrm{I}(\rbf,\omega_n) \Ebf_n(\rbf) \,\dV$ and comparing with Eq.~\eqref{eq:absorbed-power}, the absorptivity operator is found as
\begin{equation}
    \bm{\alpha}_{nm}(-\uvec{\rbf},-\uvec{\rbf}') = \sum_k \bm{\alpha}_{nmk}(-\uvec{\rbf},-\uvec{\rbf}'),
\end{equation}
where
\begin{equation}
    \bm{\alpha}_{nmk}(-\uvec{\rbf},-\uvec{\rbf}') = \frac{16\pi^2c}{\mu_0} \frac{\omega_k}{\omega_n\omega_m}  \lim_{r,r'\to\infty} rr' e^{-\im k_nr} e^{\im k_m r'}  \int_V  \mb{G}_{kn}^\dagger(\rbf'',\rbf) \bm{\varepsilon}_\mathrm{I}(\rbf'',\omega_k) \mb{G}_{km}(\rbf'',\rbf') \,\dV''
\end{equation}
is a partial absorptivity operator associated with the $k$-th frequency harmonic of the dissipation currents.

Using Eq.~\eqref{eq:reciprocity-PTMM-3}, we can show that
\begin{equation}
    \bm{\alpha}_{nmk}^\T(-\uvec{\rbf},-\uvec{\rbf}') = \frac{\omega_k^2}{\omega_n\omega_m} \bm{\epsilon}^{\mathcal{T}}_{mnk}(\uvec{\rbf}',\uvec{\rbf}). \label{eq:KLTR-op-PTVM-1}
\end{equation} 
Eq.~\eqref{eq:KLTR-op-PTVM-1} extends Kirchhoff's law of thermal radiation to the emissivity and absorptivity operators. In TIM, this equation reduces to Eq.~\eqref{eq:KLTR-op}, which, as mentioned before, has the consequence that the radiation channels of a structure are the same as the complex conjugate absorption channels of the same structure with opposite odd-symmetric bias. In PTVM, this fact applies only to the partial radiation or absorption channels associated with a specific harmonic of fluctuation or dissipation currents.


\begin{figure*}[h]
\centering
\includegraphics[width=\textwidth]{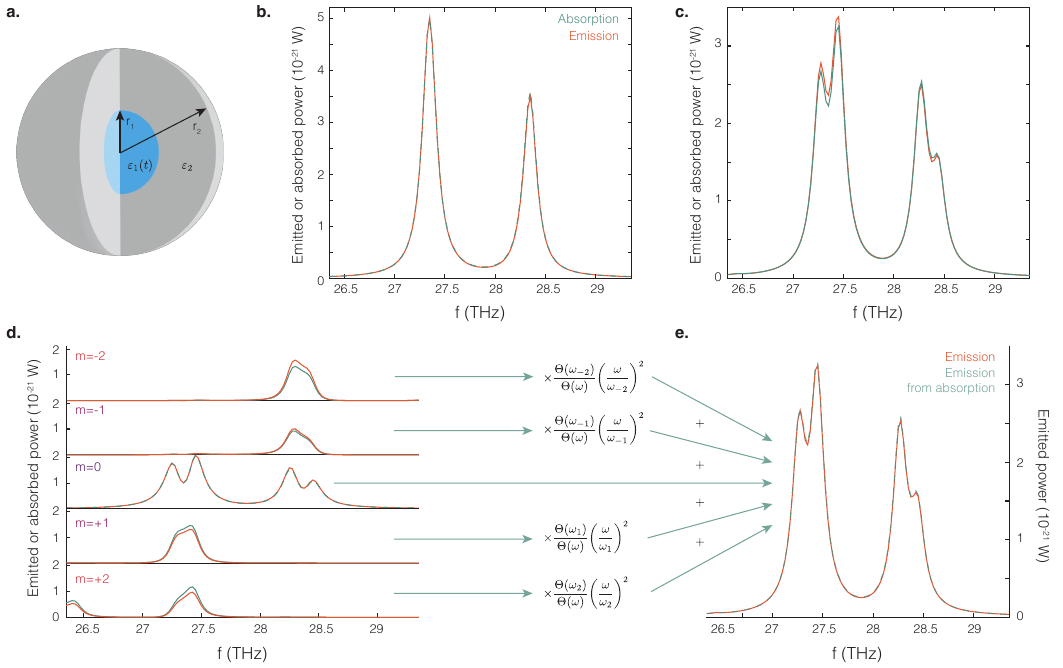}
\caption{{\bf Numerical verification of the generalized Kirchhoff's law in a full-wave Floquet system.} {\bf a}. Schematic representation of a core-shell particle with a dielectric core and a SiC shell. $r_1 = 200$ nm, $r_2 = 460$ nm, $T = 2000$ K, and $\varepsilon_1(t) = 4 + \delta \cos(2 \pi 10^{12} t)$. The permittivity of SiC is given by a permittivity model described in the text. {\bf b.} The absorbed and emitted powers of the core-shell in equilibrium with its environment \emph{without} temporal modulation ($\delta = 0$). Due to coupling between the surface-phonon polaritons on the inner and outer surfaces, two peaks appear, separated by 1 THz. {\bf c.} The absorbed and emitted powers in the presence of temporal modulation of the core permittivity with 1 THz, the difference frequency between the two peaks, with $\delta = 0.4$. A small difference between the absorbed and emitted powers is visible. {\bf d.} A decomposition of the emitted and absorbed powers into harmonics. The 0$^\text{th}$ and 1$^\text{st}$ harmonics are strong, the second harmonic is already orders of magnitude weaker. {\bf e.} Using the generalized Kirchhoff's law, we are able to reconstruct the emissivity directly from the absorptivity.}
\label{fig:S1}
\end{figure*}

\subsection{Numerical verification}\label{sm:subsec:num}\noindent
To demonstrate that the generalized Kirchhoff's law indeed holds in the full-wave case, we numerically explore absorption and emission in a core-shell particle, where the core is a (time-modulated) dielectric and the shell is silicon carbide with permittivity

\begin{equation}
\varepsilon(\omega) = \varepsilon_\infty \Biggl( 1 + \frac{\omega_L^2 - \omega_T^2}{ \omega_T^2 - \omega^2 - i \Gamma \omega} \Biggr),
\end{equation}
where $\varepsilon_\infty = 6.7$, $\omega_L = 969$ cm$^\text{-1}$, $\omega_T = 793$ cm$^\text{-1}$, and $\Gamma = 4.76$ cm$^\text{-1}$ \cite{supp-spitzer1959infrared}.

Fig.~\ref{fig:S1}a shows the core-shell particle schematically. For an inner and outed radius of 200 and 460 nm respectively, the absorption and emission spectra at of the unmodulated sphere with relative permittivity $\varepsilon_1=4$ at T=2000 K in equilibrium show two strong peaks separated by 1 THz (Fig.~\ref{fig:S1}b). These peaks arise due to coupling between the surface phonon-polaritons on the inner and outer SiC surfaces. The absorption cross section is obtained from a plane-wave excitation simulation in COMSOL, which we then multiply by Planck's law and $4\pi$ for the solid angle in order to get the total absorbed power. For the emission, we simulate single point dipoles with the radial and transverse orientation (taken into account twice) moved along a radial direction. The current density is set by the fluctuation-dissipation theorem, and for each dipole position we multiply the radiated power by the shell volume as a means to integrate the fluctuating current over the volume. The obtained emissivity and absorptivity are in excellent agreement, as expected, but calculating the emission in this way is already significantly more involved and computationally costly than calculating the absorption.

When we turn on temporal modulation of the core permitivity, $\varepsilon(t) = 4 + 0.4 \cos(\Omega t)$ with $\Omega = 2\pi 10^{12}$ rad/s the difference frequency between the two peaks, we observe splitting of each peak and a slight difference between absorptivity and emissivity (Fig.~\ref{fig:S1}c). To model a time-varying medium, we run five simulations for each frequency and incident plane wave or dipole position/orientation: the fundamental, and the first two positive and negative harmonics. The simulations are coupled via a harmonic expansion of the polarization current density. This means we need to perform five times as many simulations (for five harmonics) for the emission simulations, since currents in all harmonics result in emission at the fundamental frequency.

Similar to Fig.~2 in the main text, we can resolve the absorptivity and emissivity into its partial harmonic contributions, as shown in Fig.~\ref{fig:S1}d. By multiplying the harmonic contributions with the prefactor, we find that that the partial absorptivities can indeed be used to reconstruct the emissivity using the generalized Kirchhoff's law.

\section{Planar structure demonstrating near-complete violation of Kirchhoff's law}\label{sm:sec:planar}
\begin{figure*}[h]
\centering
\includegraphics[width=\textwidth]{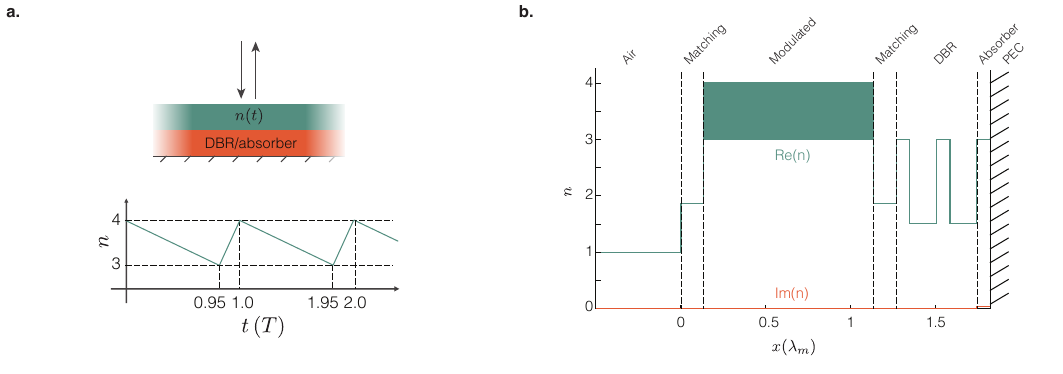}
\caption{{\bf Details on planar configuration that strongly violates the stationary Kirchhoff's law.} {\bf a}. Schematic cross section and visualization of temporal refractive index pattern, also shown in the main text. {\bf b}. Detailed representation of the multilayer structure consisting of five distinct regions: two matching layers, a modulated layer, a distributed Bragg reflector, and an absorbing layer. The structure lies on top of a perfect electric conductor and the refractive index of the absorbing layer is taken at $\omega_a$.}
\label{fig:S2}
\end{figure*}

\noindent
The planar structure of Fig.~5 in the main text consists of a time-modulated layer and a resonant cavity. The time-modulated layer has a thickness of one free-space wavelength $\lambda_m = 2\pi c/\omega_m$, and is modulated between $n_{ml} = 3$ and $n_{mh}=4$ with a sawtooth (serodyne) pattern. It is modulated with the pattern shown in Fig.~\ref{fig:S2}a, at a frequency of $\Omega = \omega_m/20$. The absorbing resonant cavity is designed for the upconverted frequency, $\omega_a = 1.05 \omega_m$, and is constructed from a layer with $n_a(\omega_a) = 3 + 0.013i$ and $d = \lambda_a/(4 \Re(n_a))$ and a distributed Bragg reflector (DBR) consisting of 4 layers with alternating high ($n_{bh} = 3$) and low ($n_{bl} = 1.5$) refractive index, each with quarter-wavelength thickness in the respective medium at $\lambda_a$. Between the DBR and the modulating layer there is a matching layer with $n_c = \sqrt{(n_{ml}+n_{mh})/2} \approx 1.87$ and $d_c = \lambda_m/(4n_c)$. The same matching layer is placed between the modulating layer and air. Fig.~\ref{fig:S2}b shows the refractive index profile.

To perform the simulations, we use a home-built 1D FDTD code. We perform narrow-band simulations and sweep the carrier frequency. Material loss is modeled through an effective conductivity $\sigma = \varepsilon_r''/\omega$ set such that $n_a(\omega_a) = 3 + 0.013i$. To perform the angle sweep, we used an effective medium approach \cite{supp-taflove2005computational}. For the TE polarization, the effective permittivity is $\varepsilon_\mathrm{eff} = \varepsilon_r - \sin^2\theta$, where $\theta$ is the incidence angle, the effective permeability is equal to $1$, and the effective conductivity is equal to the conductivity of the medium. For the TM polarization, the effective permittivity is equal to the permittivity of the medium, the effective permeability is $\mu_\mathrm{eff} = 1 - \frac{\sin^2\theta}{\varepsilon_r}$, and the effective conductivity is $\sigma_\mathrm{eff} = \sigma \frac{\sin^2\theta}{\varepsilon_r^2}$.

\newpage

\phantomsection
\label{sm:references}
\section*{Supplemental References}
\setcounter{NAT@ctr}{0}
\begin{list}{[\arabic{enumiv}]}{%
  \usecounter{enumiv}%
  \setlength{\leftmargin}{2.5em}%
  \setlength{\labelwidth}{2em}%
  \setlength{\labelsep}{0.5em}%
  \setlength{\itemsep}{0pt}%
  \setlength{\parsep}{0pt}%
}

\makeatletter
\providecommand \@ifxundefined [1]{%
 \@ifx{#1\undefined}
}%
\providecommand \@ifnum [1]{%
 \ifnum #1\expandafter \@firstoftwo
 \else \expandafter \@secondoftwo
 \fi
}%
\providecommand \@ifx [1]{%
 \ifx #1\expandafter \@firstoftwo
 \else \expandafter \@secondoftwo
 \fi
}%
\providecommand \natexlab [1]{#1}%
\providecommand \enquote  [1]{``#1''}%
\providecommand \bibnamefont  [1]{#1}%
\providecommand \bibfnamefont [1]{#1}%
\providecommand \citenamefont [1]{#1}%
\providecommand \href@noop [0]{\@secondoftwo}%
\providecommand \href [0]{\begingroup \@sanitize@url \@href}%
\providecommand \@href[1]{\@@startlink{#1}\@@href}%
\providecommand \@@href[1]{\endgroup#1\@@endlink}%
\providecommand \@sanitize@url [0]{\catcode `\\12\catcode `\$12\catcode
  `\&12\catcode `\#12\catcode `\^12\catcode `\_12\catcode `\%12\relax}%
\providecommand \@@startlink[1]{}%
\providecommand \@@endlink[0]{}%
\providecommand \url  [0]{\begingroup\@sanitize@url \@url }%
\providecommand \@url [1]{\endgroup\@href {#1}{\urlprefix }}%
\providecommand \urlprefix  [0]{URL }%
\providecommand \Eprint [0]{\href }%
\providecommand \doibase [0]{http://dx.doi.org/}%
\providecommand \selectlanguage [0]{\@gobble}%
\providecommand \bibinfo  [0]{\@secondoftwo}%
\providecommand \bibfield  [0]{\@secondoftwo}%
\providecommand \translation [1]{[#1]}%
\providecommand \BibitemOpen [0]{}%
\providecommand \bibitemStop [0]{}%
\providecommand \bibitemNoStop [0]{.\EOS\space}%
\providecommand \EOS [0]{\spacefactor3000\relax}%
\providecommand \BibitemShut  [1]{\csname bibitem#1\endcsname}%
\let\auto@bib@innerbib\@empty
\bibitem [{\citenamefont {Lifshitz}\ \emph {et~al.}(1980)\citenamefont
  {Lifshitz}, \citenamefont {Pitaevskii}, \citenamefont {Sykes},\ and\
  \citenamefont {Kearsley}}]{supp-landau-lifshitz-course}%
  \BibitemOpen
  \bibfield  {author} {\bibinfo {author} {\bibfnamefont {E.~M.}\ \bibnamefont
  {Lifshitz}}, \bibinfo {author} {\bibfnamefont {L.~P.}\ \bibnamefont
  {Pitaevskii}}, \bibinfo {author} {\bibfnamefont {J.~B.}\ \bibnamefont
  {Sykes}}, \ and\ \bibinfo {author} {\bibfnamefont {M.~J.}\ \bibnamefont
  {Kearsley}},\ }\href
  {https://app.knovel.com/hotlink/toc/id:kpLLCTPS01/landau-lifshitz-course/landau-lifshitz-course}
  {\emph {\bibinfo {title} {Landau and Lifshitz Course of Theoretical Physics,
  Statistical Physics, Part 2 - Theory of the Condensed State, Volume 9}}}\
  (\bibinfo  {publisher} {Elsevier},\ \bibinfo {year} {1980})\BibitemShut
  {NoStop}%
\bibitem [{\citenamefont {Asadchy}\ \emph {et~al.}(2020)\citenamefont
  {Asadchy}, \citenamefont {Mirmoosa}, \citenamefont {D{\'{\i}}az-Rubio},
  \citenamefont {Fan},\ and\ \citenamefont {Tretyakov}}]{supp-asadchy2020tutorial}%
  \BibitemOpen
  \bibfield  {author} {\bibinfo {author} {\bibfnamefont {V.~S.}\ \bibnamefont
  {Asadchy}}, \bibinfo {author} {\bibfnamefont {M.~S.}\ \bibnamefont
  {Mirmoosa}}, \bibinfo {author} {\bibfnamefont {A.}~\bibnamefont
  {D{\'{\i}}az-Rubio}}, \bibinfo {author} {\bibfnamefont {S.}~\bibnamefont {Fan}}, \
  and\ \bibinfo {author} {\bibfnamefont {S.~A.}\ \bibnamefont {Tretyakov}},\
  }\href {http://dx.doi.org/10.1109/JPROC.2020.3012381} {\bibfield  {journal}
  {\bibinfo  {journal} {Proceedings of the IEEE}\ }\textbf {\bibinfo {volume}
  {108}},\ \bibinfo {pages} {1684} (\bibinfo {year} {2020})}\BibitemShut
  {NoStop}%
\bibitem [{\citenamefont {Guo}\ \emph {et~al.}(2022)\citenamefont {Guo},
  \citenamefont {Zhao},\ and\ \citenamefont {Fan}}]{supp-guo2022adjoint}%
  \BibitemOpen
  \bibfield  {author} {\bibinfo {author} {\bibfnamefont {C.}~\bibnamefont
  {Guo}}, \bibinfo {author} {\bibfnamefont {B.}~\bibnamefont {Zhao}}, \ and\
  \bibinfo {author} {\bibfnamefont {S.}~\bibnamefont {Fan}},\ }\href
  {http://dx.doi.org/10.1103/PhysRevX.12.021023} {\bibfield  {journal}
  {\bibinfo  {journal} {Phys. Rev. X}\ }\textbf {\bibinfo {volume} {12}},\
  \bibinfo {pages} {021023} (\bibinfo {year} {2022})}\BibitemShut {NoStop}%
\bibitem [{\citenamefont {Miller}\ \emph {et~al.}(2017)\citenamefont {Miller},
  \citenamefont {Zhu},\ and\ \citenamefont {Fan}}]{supp-miller2017universal}%
  \BibitemOpen
  \bibfield  {author} {\bibinfo {author} {\bibfnamefont {D.~A.~B.}\
  \bibnamefont {Miller}}, \bibinfo {author} {\bibfnamefont {L.}~\bibnamefont
  {Zhu}}, \ and\ \bibinfo {author} {\bibfnamefont {S.}~\bibnamefont {Fan}},\
  }\href {\doibase 10.1073/pnas.1701606114} {\bibfield  {journal} {\bibinfo
  {journal} {Proceedings of the National Academy of Sciences of the United
  States of America}\ }\textbf {\bibinfo {volume} {114}},\ \bibinfo {pages}
  {4336} (\bibinfo {year} {2017})}\BibitemShut {NoStop}%
\bibitem [{\citenamefont {Spitzer}\ \emph {et~al.}(1959)\citenamefont
  {Spitzer}, \citenamefont {Kleinman},\ and\ \citenamefont
  {Walsh}}]{supp-spitzer1959infrared}%
  \BibitemOpen
  \bibfield  {author} {\bibinfo {author} {\bibfnamefont {W.}~\bibnamefont
  {Spitzer}}, \bibinfo {author} {\bibfnamefont {D.}~\bibnamefont {Kleinman}}, \
  and\ \bibinfo {author} {\bibfnamefont {D.}~\bibnamefont {Walsh}},\ }\bibfield
  {journal} {\bibinfo  {journal} {Physical Review}\ }\textbf {\bibinfo {volume}
  {113}},\ \bibinfo {pages} {127} (\bibinfo {year} {1959})\BibitemShut
  {NoStop}%
\bibitem [{\citenamefont {Taflove}\ \emph {et~al.}(2005)\citenamefont
  {Taflove}, \citenamefont {Hagness},\ and\ \citenamefont
  {Piket-May}}]{supp-taflove2005computational}%
  \BibitemOpen
  \bibfield  {author} {\bibinfo {author} {\bibfnamefont {A.}~\bibnamefont
  {Taflove}}, \bibinfo {author} {\bibfnamefont {S.~C.}\ \bibnamefont
  {Hagness}}, \ and\ \bibinfo {author} {\bibfnamefont {M.}~\bibnamefont
  {Piket-May}},\ }\href@noop {} {\bibfield  {journal} {\bibinfo  {journal} {The
  Electrical Engineering Handbook}\ }\textbf {\bibinfo {volume} {3}},\ \bibinfo
  {pages} {15} (\bibinfo {year} {2005})}.\BibitemShut {Stop}%

\makeatother
\end{list}

\end{document}